\documentclass[stslayout, reqno, noinfoline, preprint]{imsart}

\usepackage{soul}
\usepackage{graphicx}
\usepackage{amsthm, amssymb, amsmath}
\usepackage{epstopdf}
\usepackage{subfigure}
\usepackage{algcompatible}
\usepackage{comment}
\usepackage{tikz}
\usepackage{graphicx}

\usepackage[hmarginratio=1:1,top=32mm,columnsep=20pt]{geometry}
\usepackage{mathtools}

\usepackage[pagebackref=true, colorlinks=true, urlcolor=blue, pdfborder={0 0 0}]{hyperref}
\usepackage[sort,numbers]{natbib}
\usepackage[capitalize]{cleveref}

\startlocaldefs
\numberwithin{equation}{section}
\theoremstyle{plain}

\newtheorem{theorem}{Theorem}

\newtheorem{assumption}[theorem]{Assumptions}

\newcommand{\N}{\mathsf{N}}

\newcommand{\G}{\mathcal{G}}
\newcommand{\R}{\mathbb{R}}

\newcommand{\rd}{\mathrm{d}}

\newcommand{\bC}{\mathsf{C}}

\newcommand{\Z}{\mathbb{Z}}

\newcommand{\etgp}{\eta_{\scriptscriptstyle \textsf{GP}}}

\newcommand{\nc}{\normalcolor}
\newcommand{\Gammy}{\Gamma_{y}}
\newcommand{\Gammo}{\Gamma_{\scriptscriptstyle \textsf{obs}}}
\newcommand{\Gammgp}{\Gamma_{\scriptscriptstyle \textsf{GP}}}
\newcommand{\Gammzero}{\Gamma_{\theta}}
\newcommand{\Phit}{\Phi_{T}}

\newcommand{\Phigp}{\Phi_{\scriptscriptstyle \textsf{GP}}}
\newcommand{\Phim}{\Phi_{\scriptscriptstyle \textsf{m}}}
\newcommand{\Ces}{{({\bf C}ES)}}
\newcommand{\cEs}{{(C{\bf E}S)}}
\newcommand{\ceS}{{(CE{\bf S})}}
\newcommand{\mupost}{m}

\graphicspath{{./img/}{./img/linear/}{./img/lorenz63/}{./img/lorenz96/}{./img/darcy/}}

\begin{document}

\begin{frontmatter}
		\title{Calibrate, Emulate, Sample}
		\runtitle{}
		\begin{aug}
				\author{\fnms{Emmet} \snm{Cleary}\ead[label=e1]{ecleary@caltech.edu}},
				\author{\fnms{Alfredo} \snm{Garbuno-Inigo}\ead[label=e2]{agarbuno@caltech.edu}},
				\author{\fnms{Shiwei} \snm{Lan}\ead[label=e3]{slan@asu.edu}},
				\author{\fnms{Tapio} \snm{Schneider}\ead[label=e4]{tapio@caltech.edu}}
				\and
				\author{\fnms{Andrew M.} \snm{Stuart}\ead[label=e5]{astuart@caltech.edu}}
				\runauthor{Cleary, Garbuno-Inigo, Lan, Schneider \& Stuart}

				\address[cal]{California Institute of Technology, Pasadena, CA.
				\printead*{e1,e2,e4,e5}
				}

				\address[ariz]{Arizona State University, Tempe, AZ.
        \printead*{e3}
        }
		\end{aug}

		\maketitle
		\begin{abstract}
				Many parameter estimation problems arising in applications are best cast
				in the framework of Bayesian inversion. This allows not only for an
				estimate of the parameters, but also for the quantification of
				uncertainties in the estimates. Often in such problems the
				parameter-to-data map is very expensive to evaluate, and computing
				derivatives of the map, or derivative-adjoints, may not be feasible.
				Additionally, in many applications only noisy evaluations of the map may
				be available. We propose an approach to Bayesian inversion in such
				settings that builds on the derivative-free optimization capabilities of
				ensemble Kalman inversion methods. The overarching approach is to first
				use ensemble Kalman sampling (EKS) to {\em calibrate} the unknown
				parameters to fit the data; second, to use the output of the EKS to {\em
				emulate} the parameter-to-data map; third, to {\em sample} from an
				approximate Bayesian posterior distribution in which the
				parameter-to-data map is replaced by its emulator. This results in a
				principled approach to approximate Bayesian inference that requires only
				a small number of evaluations of the (possibly noisy approximation of
				the) parameter-to-data map. It does not require derivatives of this map,
				but instead leverages the documented power of ensemble Kalman methods.
				Furthermore, the EKS has the desirable property that it evolves the
				parameter ensembles towards the regions in which the bulk of the
				parameter posterior mass is located, thereby locating them well for the
				emulation phase of the methodology. In essence, the EKS methodology
				provides a cheap solution to the design problem of where to
				place points in parameter space to efficiently train an emulator of the
				parameter-to-data map for the purposes of Bayesian inversion.

		\end{abstract}
\end{frontmatter}


\section{Introduction}
\label{sec:I}


Ensemble Kalman methods have proven to be highly successful for state estimation
in noisily observed dynamical systems
\citep{kalnay2003atmospheric,evensen2009data,majda2012filtering,abarbanel2013predicting,asch2016data,Cotter2013,law2015data,Houtekamer16a}.
They are widely used, especially within the geophysical sciences and numerical
weather prediction, because the methodology is derivative-free, provides
reliable state estimation with a small number of ensemble members, and, through
the ensemble, provides information about sensitivities. The empirical success in
state estimation has led to further development of ensemble Kalman methods in
the solution of inverse problems, where the objective is the estimation of
parameters rather than states. Its use as an iterative method for parameter
estimation originates in the papers \citep{chen2012ensemble,emerick2013ensemble}
and recent contributions are discussed in
\citep{albers2019ensemble,evensen2018analysis,iglesias2013ensemble,stuart2018inverse}.
But despite their widespread use for both state and parameter estimation,
ensemble Kalman methods do not provide a basis for systematic uncertainty
quantification, except in the Gaussian case
\citep{le2009large,ernst2015analysis}. This is for two primary reasons: (i) the
methods invoke a Gaussian ansatz, which is not always justified; (ii) they are
often employed in situations where evaluation of the underlying dynamical system
(state estimation) or forward model (parameter estimation) is very expensive,
and only a small ensemble is feasible. The goal of this paper is to develop a
method that provides a basis for systematic Bayesian uncertainty quantification
within inverse problems, building on the proven power of ensemble Kalman
methods. The basic idea is simple: we \emph{calibrate} the model using variants
of ensemble Kalman inversion; we use the evaluations of the forward model made
during the calibration to train an \emph{emulator}; we perform approximate
Bayesian inversion using Markov Chain Monte Carlo (MCMC) \emph{samples} based on
the (cheap) emulator rather than the original (expensive) forward model. Within
this overall strategy, the ensemble Kalman methods may be viewed as providing a
cheap and effective way of determining an experimental design for training an
emulator of the parameter-to-data map to be employed within MCMC-based Bayesian
parameter estimation; this is the primary innovation contained within the paper.

\subsection{Literature Review}

The ensemble Kalman approach to calibrating unknown parameters to data is
reviewied in \citep{oliver2008inverse,iglesias2013ensemble}, and the imposition
of constraints within the methodology is overviewed in
\citep{albers2019ensemble}. We refer to this class of methods for calibration,
collectively, as ensemble Kalman inversion (EKI) methods and note that
pseudo-code for a variety of the methods may be found in
\citep{albers2019ensemble}. An approach to using ensemble-based methods to
produce approximate samples from the Bayesian posterior distribution on the
unknown parameters is described in \citep{garbuno2019gradient}; we refer to this
method as ensemble Kalman sampling (EKS). Either of these approaches, EKI or
EKS, may be used in the calibration step of our approximate Bayesian inversion
method.

Gaussian processes (GPs) have been widely used as emulation tools for
computationally expensive computer codes \citep{Santner2013}. The first use of
GPs in the context of uncertainty quantification was proposed in modeling ore
reserves in mining \citep{Krige1951}. Its motivation was a method to find the
best linear unbiased predictor, known as kriging in the geostatistics community
\citep{Cressie1993, Stein2012}. It was later adopted in the field of computer
experiments \citep{Sacks1989} to model possibly correlated residuals. The idea
was then incorporated within a Bayesian modeling perspective \citep{Currin1991}
and has been gradually refined over the years. Kennedy and O'Hagan
\citep{Kennedy2001} offer a mature perspective on the use of GPs as emulators,
adopting a clarifying Bayesian formulation. The use of GP emulators covers a
wide range of applications such as uncertainty analysis \citep{Oakley2002},
sensitivity analysis \citep{Oakley2004}, and computer code calibration
\citep{Higdon2004}. Perturbation results for the posterior distribution, when
the forward model is approximated by a GP, may be found in
\citep{stuart2018posterior}. We will exploit GPs for the \emph{emulation} step of
our method which, when informed by the calibration step, provides a robust
approximate forward model. Neural networks \citep{Goodfellow-et-al-2016} could
also be used in the emulation step and may be preferable for some applications
of the proposed methodology.

Bayesian inference is now widespread in many areas of science and engineering,
in part because of the development of generally applicable and easily
implementable sampling methods for complex modeling scenarios. MCMC methods
\citep{Metropolis1949,Metropolis1953,Hastings1970} and sequential Monte Carlo
(SMC) \citep{del2006sequential} provide the primary examples of such methods,
and their practical success underpins the widespread interest in the Bayesian
solution of inverse problems \citep{kaipio2006statistical}. We will employ MCMC
for the \emph{sampling} step of our method. SMC could equally well be used for
the sampling step and will be preferable for many problems; however, it is a
less mature method and typically requires more problem-specific tuning than
MCMC.

The impetus for the development of our approximate Bayesian inversion method is
the desire to perform Bayesian inversion on computer models that are very
expensive to evaluate, for which derivatives and adjoint calculations are not
readily available and are possibly noisy. Ensemble Kalman inversion methods
provide good parameter estimates even with many parameters, typically with
$\mathcal{O}(10^2)$ forward model evaluations
\citep{kalnay2003atmospheric,oliver2008inverse}, but without systematic
uncertainty quantification. While MCMC and SMC methods provide systematic
uncertainty quantification the fact that they require many iterations, and hence
evaluations of the forward model in our setting, is well-documented
\citep{Geyer2002}. Several diagnostics for MCMC convergence are available
\citep{Robert2004}, and theoretical guarantees of convergence exist
\citep{Meyn2012}. The rate of convergence for MCMC is determined by the size of
the step arising from proposal distribution: short steps are computationally
inefficient as the parameter space is only locally explored whereas large steps
lead to frequent rejections and hence to a waste of computational resources (in
our setting forward model evaluations) to generate additional samples. In
practice MCMC often requires of $\mathcal{O}(10^5)$ forward model evaluations
\citep{Geyer2002}. This is not feasible, for example, with climate models
\citep{Emmet,jarvinen2012ensemble,schneider2017earth}.

In the sampling step of our approximate Bayesian inversion method, we use an
emulator that can be evaluated rapidly in place of the computationally expensive
forward model, leading to Bayesian parameter estimation and uncertainty
quantification essentially at the computational cost of ensemble Kalman
inversion. The ensemble methods provide an effective design for the emulation,
which makes this cheap cost feasible.


In some applications, the dimension of the unknown parameter is high and it is
therefore of interest to understand how the constituent parts of the proposed
methodology behave in this setting. The growing use of ensemble methods
reflects, in part, the empirical fact that they scale well to high-dimensional
state and parameter spaces, as demonstrated by applications in the geophysical
sciences \citep{kalnay2003atmospheric, oliver2008inverse}; thus, the calibrate
phase of the methodology scales well with respect to high input dimension.
Gaussian process regression does not, in general, scale well to high-dimensional
input variables, but alternative emulators, such as those based on neural
networks \citep{Goodfellow-et-al-2016} are empirically found to do so; thus, the
emulate phase can potentially be developed to scale well with respect to high
input dimensions. Standard MCMC methods do not scale well with respect to high
dimensions; see \citep{Roberts2004} in the context of i.i.d. random variables in
high dimensions. However the reviews \citep{Cotter2013,chen2018robust} describe
non-traditional MCMC methods which overcome these poor scaling results for
high-dimensional Bayesian inversion with Gaussian priors, or transformations of
Gaussians; the paper \citep{kantas2014sequential} builds on the ideas in
\citep{Cotter2013} to develop SMC methods that are efficient in high- and
infinite-dimensional spaces. Thus, the calibrate phase of the methodology
scales well with respect to high input dimension for appropriately chosen
priors.


\subsection{Our Contribution}
\label{ssec:OC}

\begin{itemize}

\item We introduce a practical methodology for approximate Bayesian parameter
learning in settings where the parameter-to-data map is expensive to evaluate,
not easy to differentiate or not differentiable, and where evaluations are
possibly polluted by noise. The methodology is modular and broken into three
steps, each of which can be tackled by different methodologies:
\emph{calibration, emulation, and sampling.}

\item In the {\em calibration} phase we leverage the power of ensemble Kalman
methods, which may be viewed as fast derivative-free optimizers or approximate
samplers. These methods provide a cheap solution to the experimental design
problem and ensure that the forward map evaluations are well-adapted to the task
of Gaussian process \emph{emulation}, within the context of an outer Bayesian
inversion loop via MCMC \emph{sampling}.

\item We also show that, for problems in which the forward model evaluation is
inherently noisy, the Gaussian process emulation serves to remove the noise,
resulting in a more practical Bayesian inference via MCMC.

\item We demonstrate the methodology with numerical experiments on a model
linear problem, on a Darcy flow inverse problem, and on the Lorenz '63 and '96
models.
\end{itemize}

In \cref{sec:M}, we describe the calibrate-emulate-sample methodology introduced
in this paper, and in \cref{sec:L}, we demonstrate the method on a linear
inverse problem whose Bayesian posterior is explicitly known.
In \cref{sec:D}, we study the inverse problem of determining permeability from
pressure in Darcy flow, a nonlinear inverse problem in which the coefficients of
a linear elliptic partial differential equation (PDE) are to be determined from
linear functionals of its solution. \cref{sec:T} is devoted to the inverse
problem of determining parameters appearing in time-dependent differential
equations from time-averaged functionals of the solution. We view finite
time-averaged data as noisy infinite time-averaged data and use GP emulation to
estimate the parameter-to-data map and the noise induced through finite-time
averaging. Applications to the Lorenz '63 and '96 atmospheric science models are
described here, and use of the methodology to study an atmospheric general
circulation model is described in the paper \citep{Emmet}.

\section{Calibrate-Emulate-Sample}
\label{sec:M}


\begin{figure}[!htp]
	\includegraphics[width=\linewidth]{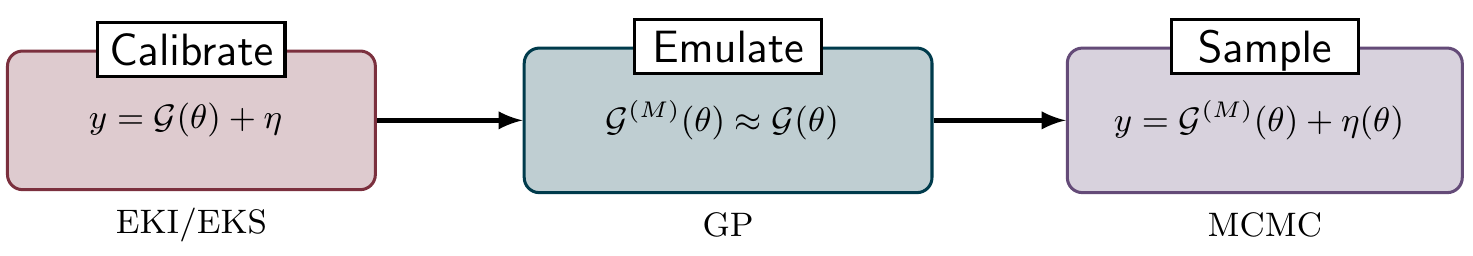}
	\caption{Schematic of approximate Bayesian inversion method to find $\theta$
	from $y$. EKI/EKS produce a small number of approximate (expensive) samples
	$\{\theta^{(m)}\}_{m=1}^M.$ These are used to train a GP approximation
	$\G^{(M)}$ of $\G$, used within MCMC to produce a large number of approximate
	(cheap) samples $\{\theta^{(n)}\}_{n=1}^{N_s}$, $N_s \gg M.$}
	\label{fig:ces-framework}
\end{figure}

\subsection{Overview}

Consider parameters $\theta$ related to data $y$ through the forward
model $\G$ and noise $\eta$:
\begin{align}
\label{eq:IP}
		y = \G(\theta) + \eta.
\end{align}
The inverse problem is to find unknown $\theta$ from $y$, given knowledge of
$\G:\R^p\to\R^d$ and some information about the noise level such as its size
(classical approach) or distribution (statistical approach), but not its value.
To formulate the Bayesian inverse problem, we assume, for simplicity, that the
noise is drawn from a Gaussian with distribution $N(0,\Gammy)$, that the prior
on $\theta$ is a Gaussian $N(0,\Gammzero)$, and that $\theta$ and $\eta$ are
{\em a priori} independent. If we define\footnote{For any positive-definite
symmetric matrix $A$, we define $\langle a,a' \rangle_{A}=\langle a,A^{-1}a'
\rangle= \langle A^{-\frac12}a,A^{-\frac12}a' \rangle$ and
$\|a\|_{A}=\|A^{-\frac12}a\|.$}
\begin{equation}\label{eq:phi}
		\Phi_R(\theta)=\frac{1}{2}\|y-\G(\theta)\|^2_{\Gammy}+\frac{1}{2}\|\theta\|^2_{\Gammzero},
\end{equation}
the posterior on $\theta$ given $y$ has density
\begin{equation}
\label{eq:post2}
 \pi^y(\theta) \propto \exp\bigl(-\Phi_R(\theta)\bigr).
\end{equation}

In a class of applications of particular interest to us, the data $y$ comprises
statistical averages of observables. The map $\G(\theta)$ provides the
corresponding statistics delivered by a model that depends on $\theta.$ In this
setting the assumption that the noise be Gaussian is reasonable if $y$ and
$\G(\theta)$ represent statistical aggregates of quantities that vary in space
and/or time. Additionally, we take the view that parameters $\theta'$ for which
Gaussian priors are not appropriate (for example because they are constrained to
be positive) can be transformed to parameters $\theta$ for which Gaussian priors
make sense.

We use EKS with $J$ ensemble members and $N$ iterations (time-steps of a
discretized continuous time algorithm)  to generate approximate samples from
\eqref{eq:post2}, in the \emph{calibration} step of the methodology. This gives
us $JN$ parameter--model evaluation pairs $\{\theta^{(m)},
\G(\theta^{(m)})\}_{m=1}^{JN}$ which we can use to produce an approximation of
$\G$ in the GP \emph{emulation} step of the algorithm. Whilst the methodology of
optimal experimental design can be used, in principle, as the basis for choosing
parameter-data pairs for the purpose of emulation \citep{alexanderian2016fast}
it can be prohibitively expensive. The theory and numerical results shown in
\citep{garbuno2019gradient} demonstrate that EKS distributes particles in regions
of high posterior probability; this is because it approximates a mean-field
interacting particle system with invariant measure equal to the posterior
probability distribution \eqref{eq:post2}. Using the EKS thus provides a cheap
and effective solution to the design problem, producing parameter--model
evaluation pairs that are well-positioned for the task of approximate Bayesian
inversion based on the (cheap) emulator. In practice it is not always necessary
to use all $JN$  parameter--model evaluation pairs but to instead use a subset
of size $M \le JN$; we denote the  resulting GP approximation of $\G$ by
$\G^{(M)}$. Throughout this paper we simply take $M=J$ and use the output of the
EKS in the last step of the iteration as the design. However other strategies,
such as descreasing $J$ and using all or most of the $N$ steps, are also
feasible.

$\G^{(M)}$ in place of the (expensive) forward model
$\G$. With the emulator $\G^{(M)}$, we define the modified inverse problem of
finding parameter $\theta$ from data $y$ when they are assumed to be related,
through noise $\eta$, by
\begin{align*}
		y = \G^{(M)}(\theta) + \eta.
\end{align*}
This is an approximation of the inverse problem defined by \eqref{eq:IP}. The
approximate posterior on $\theta$ given $y$ has density $\pi^{(M)}$ defined by
the approximate log likelihood arising from this approximate inverse problem. In
the \emph{sample} step of the methodology, we apply MCMC methods to sample from
$\pi^{(M)}.$

The overall framework, comprising the three steps calibrate, emulate, and sample
is cheap to implement because it involves a small number of evaluations of
$\G(\cdot)$ (computed only during the calibration phase, using ensemble methods
where no derivatives of $\G(\cdot)$ are required), and because the MCMC method,
which may require many steps, only requires evaluation of the emulator
$\G^{(M)}(\cdot)$ (which is cheap to evaluate) and not $\G(\cdot)$ (which is
assumed to be costly to evaluate and is possibly only known noisily).
On the other hand, the method has ingredients
that make it amenable to produce a controlled approximation of the true
posterior. This is true since the EKS steps generate points concentrating near
the main support of the posterior so that the GP emulator provides an accurate
approximation of $\G(\cdot)$ where it matters.\footnote{By ``main support'' we
mean a region containing the majority of the probability.} A depiction of the
framework and the algorithms involved can be found in \cref{fig:ces-framework}.
In the rest of the paper, we use the acronym CES to denote the three step
methodology. Furthermore we employ boldface on one letter when we wish to
emphasize one of the three steps: calibration is emphasized by \Ces; emulation
by \cEs; and sampling by \ceS.

\subsection{Calibrate -- EKI And EKS}\label{ssec:calibrate}

The use and benefits of ensemble Kalman methods to solve inverse or parameter
calibration problems have been outlined in the introduction. We will employ
particular forms of the ensemble Kalman inversion methodology that we have found
to perform well in practice and that are amenable to analysis; however, other
ensemble methods to solve inverse problems could be used in the calibration
phase.

The basic EKI method is found by time-discretizing the following system of
interacting particles \citep{schillings2017analysis}:
\begin{align} \label{eq:eki-implement}
\frac{d \theta^{(j)}}{dt} = - \frac{1}{J}\sum_{k = 1}^J \, \langle \G(\theta^{(k)}) - \bar{\G}, \G(\theta^{(j)}) - y \rangle_{\Gammy} \, (\theta^{(k)}-\bar{\theta}),
\end{align}
where $\bar{\theta}$ and $\bar{\G}$ denote the sample means given by
\begin{align}
 \bar{\theta} &= \frac{1}{J} \sum_{k = 1}^J \theta^{(k)}, \qquad
\bar{\G} = \frac{1}{J} \sum_{k = 1}^J \G(\theta^{(k)}).
\label{eq:SM}
\end{align}
For use below, we also define $\Theta=\{\theta^{(j)}\}_{j=1}^J$ and the $p
\times p$ matrix
\begin{align}
 \bC(\Theta) &= \frac{1}{J} \sum_{k = 1}^J (\theta^{(k)} - \bar{\theta} ) \otimes
(\theta^{(k)} - \bar{\theta}).
\label{eq:CC}
\end{align}
The dynamical system \eqref{eq:eki-implement} drives the particles to consensus,
while also driving them to fit the data and hence solve the inverse problem
\eqref{eq:IP}. Time-discretizing the dynamical system leads to a form of
derivative-free optimization to minimize the least squares misfit defined by
\eqref{eq:IP} \citep{iglesias2013ensemble,schillings2017analysis}.

An appropriate modification of EKI, to attack the problem of sampling from the
posterior $\pi^y$ given by \eqref{eq:post2}, is EKS \citep{garbuno2019gradient}.
Formally this is obtained by adding a prior-related damping term, as in
\citep{chada2019tikhonov}, and a $\Theta$-dependent noise to obtain
\begin{align} \label{eq:implement}
\frac{d\theta^{(j)}}{dt} = - \frac{1}{J}\sum_{k = 1}^J \, \langle \G(\theta^{(k)}) - \bar{\G}, \G(\theta^{(j)}) - y \rangle_{\Gammy} \, (\theta^{(k)}-\bar{\theta}) \, - \,
\bC(\Theta) \Gammzero^{-1}\theta^{(j)} +\,\sqrt{2\bC(\Theta)} \, \frac{d{W}^{(j)}}{dt},
\end{align}
where the $\{W^{(j)}\}$ are a collection of i.i.d. standard Brownian motions in
the parameter space $\R^p.$ The resulting interacting particle system
approximates a mean-field Langevin-McKean diffusion process which, for linear
$\mathcal{G}$, is invariant with respect to the posterior distribution
\eqref{eq:post2} and, more generally, concentrates close to it; see
\citep{garbuno2019gradient} and \citep{garbuno2019affine} for details. The
specific algorithm that we implement here time-discretizes \eqref{eq:implement}
by means of a linearly implicit split-step scheme given by
\citep{garbuno2019gradient}
\begin{subequations}\label{eq:implicit}
\begin{align}
		{\theta}^{(*, j)}_{n+1} &= {\theta}^{(j)}_{n} - \Delta t_n \, \frac{1}{J}\sum_{k = 1}^J \, \langle \G(\theta^{(k)}_n) - \bar{\G}, \G(\theta^{(j)}_n) - y \rangle_{\Gammy} \, \theta^{(k)}_n \, - \Delta t_n \, \bC (\Theta_n) \, \Gammzero^{-1} \, {\theta}^{(*, j)}_{n+1} \\
		{\theta}^{(j)}_{n+1} &= {\theta}^{(*, j)}_{n+1} +\,\sqrt{2 \, \Delta t_n\, \bC(\Theta_n)} \, \xi^{(j)}_n,
\end{align}
\end{subequations}
where $\xi^{(j)}_n \sim \N(0, I)$, $\Gammzero$ is the prior covariance and
$\Delta t_n$ is an adaptive timestep give in \citep{garbuno2019gradient}, and
based on methods developed for EKI in \citep{kovachki2019ensemble}. The
finite $J$ correction to \eqref{eq:implement} proposed in \citep{nusken2019note},
and further developed in \citep{garbuno2019affine}, can easily be incorporated
into the explicit step of this algorithm, and other time-stepping methods can
also be used.

\subsection{Emulate -- GP Emulation}\label{ssec:emulate}

The ensemble-based algorithm described in the preceding subsection produces
input-output pairs $\{\theta^{(i)}_n, \G(\theta^{(i)}_n)\}_{i=1}^J$ for $n=0,
\dots, N$. For $n=N$ and $J$ large enough, the samples of $\theta$ are
approximately drawn from the posterior distribution. We use a subset of
cardinality $M \leq JN$ of this design as training points to update a GP prior
to obtain the function $\G^{(M)}(\cdot)$ that will be used instead of the true
forward model $\G(\cdot)$. The cardinality $M$ denotes the total number of
evaluations of $\G$ used in training the emulator $\G^{(M)}$. Recall that
throughout this paper we take $M=J$ and use the output of the
EKS in the last step of the iteration as the design.

The forward model is a multioutput map $\G: \R^p \mapsto \R^d.$ It often
suffices to emulate each output coordinate $l = 1, \ldots, d$ independently;
however, variants on this are possible, and often needed, as discussed at the end
of this subsection. For the moment, let us consider the emulation of the $l$-th
component in $\G(\theta)$; denoted by $\G_l(\theta)$. Rather than interpolate
the data, we assume that the input-output pairs are polluted by additive
noise.\footnote{The paper \citep{andrianakis2012} interprets the use of additive
noise within computer code emulation.} We place a Gaussian process prior with
zero or linear mean function on the $l$-th output of the forward model and, for
example, use the squared exponential kernel
\begin{align}
		k_l(\theta, \theta') = \sigma^2_l \, \exp\left( -\frac12 \|\theta - \theta' \|^2_{D_l} \right) + \lambda^2_l \, \delta_{\theta}(\theta'),\label{eq:gp_rbf}
\end{align}
where $\sigma_l$ denotes the amplitude of the covariance kernel; $\sqrt{D_l} =
\text{diag}(\ell_1^{(l)}, \ldots, \ell_p^{(l)})$, is the diagonal matrix of
lengthscale parameters; $\delta_x(y)$ is the Kronecker delta function; and
$\lambda_l$ the standard deviation of a homogenous white noise capturing the
assumed noise in the input-output pairs. The hyperparameters $\phi_l =
(\sigma^2_l, D_l, \lambda^2_l)$, which are learnt from the input-output pairs
along with the regression coefficients of the GP, account for signal strength
$(\sigma^2_l);$ sensitivity to changes in each parameter component $(D_l);$ and
the possibility of white noise with variance $\lambda^2_l$ in the evaluation of
the $l$-th component of the forward model $\G_l(\cdot).$ We adopt an empirical
Bayes approach to learn the hyperparameters of each of the $d$ Gaussian
processes.

The final emulator is formed by stacking each of the GP models in a vector,
\begin{align}\label{eq:gp-emulator-noisy}
\G^{(M)}(\theta) \sim \N \left(m(\theta), \Gammgp(\theta)\right).
\end{align}
The noise $\eta$ typically found in \eqref{eq:IP} needs to be incorporated along
with the noise $\etgp(\theta)\sim \N \left(0, \Gammgp(\theta)\right)$ in the
emulator $\G^{(M)}(\theta)$, resulting in the inverse problem
\begin{align}
\label{eq:IP2}
		y = m(\theta) + \etgp(\theta)+\eta.
\end{align}

We assume that $\etgp(\theta)$ and $ \eta$ are independent of one another. In
some cases, one or other of the sources of noise appearing in \eqref{eq:IP2} may
dominate the other and we will then neglect the smaller one. If we neglect
$\eta,$ we obtain the negative log-likelihood
\begin{align}
\label{eq:phi_gp}
	\Phigp^{(M)}(\theta) = \frac{1}{2} \|y-m(\theta)\|^2_{\Gammgp(\theta)} + \frac12 \log \det\Gammgp(\theta);
\end{align}
for example, in situations where initial conditions of a dynamical systems are
not known exactly, as encountered in time-average applications {\cref{sec:T}}
the noise $\etgp(\theta)$ is deemed to be the major source of uncertainty and we
take $\eta=0.$ If, on the other hand, we neglect $\etgp$ then we obtain negative
log-likelihood
\begin{align}
\label{eq:phi_m}
	\Phim^{(M)}(\theta) = \frac{1}{2} \|y-m(\theta)\|^2_{\Gammy}.
\end{align}
If both noises are incorporated, then \eqref{eq:phi_gp} is modified to give
\begin{align}
\label{eq:phi_gp2}
	\Phigp^{(M)}(\theta) = \frac{1}{2} \|y-m(\theta)\|^2_{\Gammgp(\theta)+\Gammy} + \frac12 \log \det\bigl(\Gammgp(\theta)+\Gammy\bigr);
\end{align}
This is used in \cref{sec:D}.

We note that the specifics of the GP emulation could be adapted -- we could use
correlations in the output space $\R^d$, other kernels, other mean functions,
and so forth. A review of multioutput emulation in the context of machine
learning can be found in \citep{alvarez2012} and references therein; specifics
on decorrelating multioutput coordinates can be found in \citep{higdon2008}; and
recent advances in exploiting covariance structures for multioutput emulation in
\citep{bilionis2013,atkinson2019}. For the sake of simplicity, we will focus on
emulation techniques that preserve the strategy of determining, and then
learning, approximately uncorrelated components; methodologies for transforming
variables to achieve this are discussed in \cref{apx:diag}.

\subsection{Sample -- MCMC}\label{ssec:sample}

For the purposes of the MCMC algorithm, we need to initialize the Markov chain
and choose a proposal distribution. The Markov chain is initialized with
$\theta_0$ drawn from the support of the posterior; this ensures that the MCMC
has a short transient phase. To initialize, we use the ensemble mean of the
EKS at the last iteration, $\bar \theta_J$. For the MCMC step we use a proposal
of random walk Metropolis type, employing a multivariate Gaussian distribution
with covariance given by the empirical covariance of the ensemble from EKS. We
are thus pre-conditioning the sampling phase of the algorithm with approximate
information about the posterior from the EKS. The resulting MCMC method is
summarized in the following steps and is iterated until a desired number $N_s
\gg J$ of samples $\{\theta_n\}_{n=1}^{N_s}$ is generated:
\begin{enumerate}
		\item Choose $\theta_0=\bar \theta_J$.
		\item Propose a new parameter choice $\theta^*_{n+1}=\theta_n + \xi_n$ where $\xi_n
				\sim N(0,C(\Theta_J))$.
		\item Set $\theta_{n+1}=\theta_{n+1}^{*}$ with probability $a(\theta_{n},\theta^*_{n+1})$;
				otherwise set $\theta_{n+1} = \theta_n$.
		\item $n \to n+1$, return to 2.
\end{enumerate}
The acceptance probability is computed as:
\begin{align}
a(\theta,\theta^*) = \min \left\lbrace1, \exp\left[\left(\Phi_{\cdot}^{(M)}(\theta^*) + \frac12 \|\theta^*\|^2_{\Gammzero}\right) - \left(\Phi_{\cdot}^{(M)}(\theta) + \frac12 \|\theta\|^2_{\Gammzero}\right) \right]\right\rbrace,
\end{align}
where $\Phi_{\cdot}^{(M)}$ is defined in \eqref{eq:phi_gp}, \eqref{eq:phi_m} or \eqref{eq:phi_gp2}, whichever is appropriate. \\

\section{Linear Problem}
\label{sec:L}

By choosing a linear parameter-to-data map, we illustrate the methodology in a
case where the posterior is Gaussian and known explicitly. This demonstrates
both the viability and accuracy of the method in a transparent fashion.

\subsection{Linear Inverse Problem}
\label{ssec:TSD1}

We consider a linear forward map $\G(\theta) = G \theta$, with $G \in \R^{d
\times p}$. Each row of the matrix $G$ is a $p$-dimensional draw from a
multivariate Gaussian distribution. Concretely we take $p=2$ and each row $G_i
\sim \N(0, \Sigma)$, where $\Sigma_{12} = \Sigma_{21} = -0.9$, and $\Sigma_{11}
= \Sigma_{22} = 1.$ The synthetic data we have available to perform the Bayesian
inversion is then given by
\begin{align}
	y = G\theta^\dagger + \eta,
\end{align}
where $\theta^{\dagger} = [-1, 2]^\top$, and $\eta \sim \N(0,\Gamma)$ with
$\Gamma= 0.1^2 I$.

We assume that, a priori, parameter $\theta \sim
\N(\mupost_{\theta},\Sigma_{\theta})$. In this linear Gaussian setting the
solution of the Bayesian linear inverse problem is itself Gaussian
\cite[see][Part IV]{gelman2013bayesian} and given by the Gaussian distribution
\begin{align}
	\pi^y(\theta) \propto \exp\left(-\frac12 \| \theta - \mupost_{\theta|y}\|^2_{\Sigma_{\theta|y}}\right),
\end{align}
where $\mupost_{\theta|y}$ and $\Sigma_{\theta|y}$ denote the posterior mean and
covariance. These are computed as
\begin{align}
	\Sigma_{\theta|y}^{-1} = G^\top \Sigma^{-1}_y G + \Sigma_{\theta}^{-1}, \qquad \mupost_{\theta|y} = \Sigma_{\theta|y} \left( G^\top \Sigma^{-1}_y y + \Sigma_{\theta}^{-1} \mupost_{\theta}\right).
\end{align}
\subsection{Numerical Results}
\label{ssec:LR}

\begin{figure}[!ht]
	\centering
	\includegraphics[width=.99\linewidth]{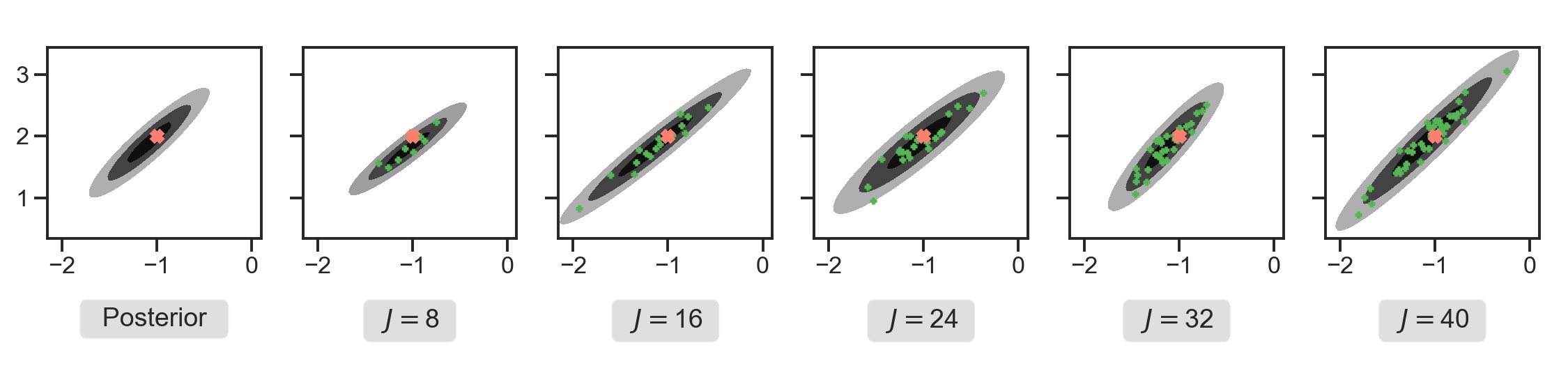}
	\caption{Density estimates for different ensemble sizes used for the
	calibration step. Leftmost panel show the true posterior distribution. The
	green dots show the EKS at the $20$-th iteration. The contour levels show the
	density of a Gaussian with mean and covariance estimated from EKS at said
	iteration. }
	\label{fig:lin-calibrate}
\end{figure}

For the calibration step \Ces\ we consider the EKS algorithm.
Figure~\ref{fig:lin-calibrate} shows how the EKS samples estimate the posterior
distribution (the far left). The green dots correspond to $ 20$-th iteration of
EKS with different ensemble sizes. We also display in gray contour levels the
density corresponding to the $67\%,$ $ 90\%$ and $99\%$ probability levels under
a Gaussian with mean and covariance estimated from EKS at said $20$-th
iteration. This allows us to visualize the difference between the results of EKS
and the true posterior distribution in the leftmost panel. In this linear case,
the mean-field limit of EKS exactly reproduces the invariant measure
\citep{garbuno2019gradient}. The mismatch between the EKS samples and the true
posterior can be understood from the fact that time discretizations of Langevin
diffusions are known to induce errors if no metropolization scheme is added to
the dynamics
\citep{leimkuhler2018ensemble,roberts1998optimal,roberts1996exponential}, and
from the finite number of particles used; the latter could be corrected by using
the ideas introduced in \citep{nusken2019note} and further developed in
\citep{garbuno2019affine}.

\begin{figure}[!ht]
	\centering
	\includegraphics[width=.95\linewidth]{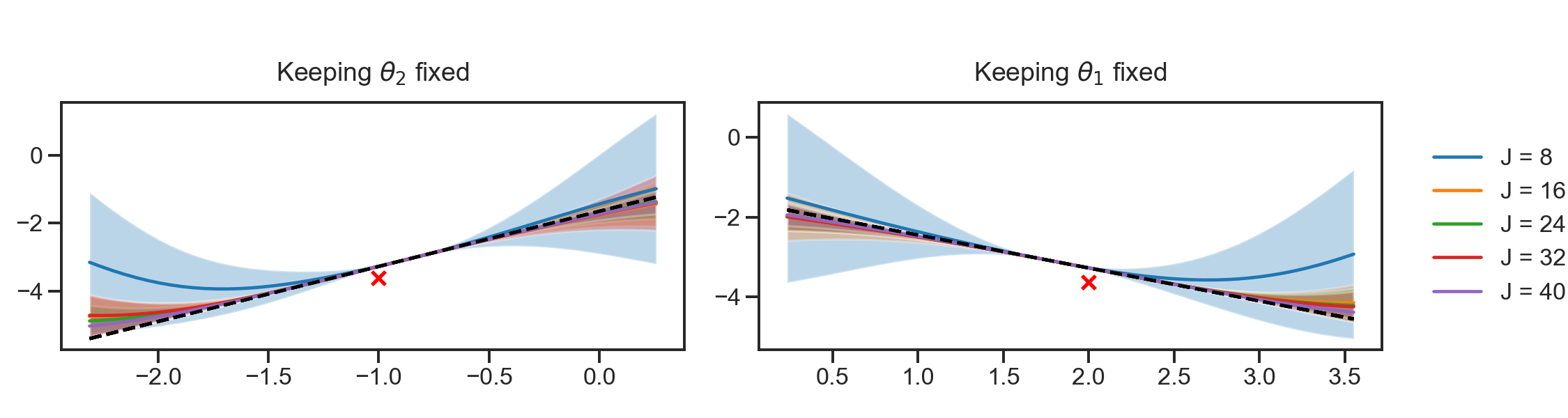}
	%
	\caption{Gaussian process emulators learnt using different training datasets.
	The input-output pairs are obtained from the calibration step using EKS.}
	\label{fig:lin-emulate}
\end{figure}

The GP emulation step \cEs\ is depicted in \cref{fig:lin-emulate} for one
component of $\G$. Each GP is trained using the $20$-th iteration of EKS for
each ensemble size. We employ a zero mean GP with the squared-exponential kernel
\eqref{eq:gp_rbf}. We add a regularization term to the lengthscales of the GP by
means of a prior in the form of a Gamma distribution. This is common in GP
Bayesian inference when the domain of the input variables is unbounded. The
choice of this regularization ensures that the covariance kernels, which are
regression functions for the mean of the emulator, decay fast away from the
data, and that no short variations below the levels of available data are
introduced \citep{gelman2017prior,gelman2013bayesian}. We can visualize the
emulator of the component of the linear system considered here by fixing one
parameter at the true value while varying the other. The dashed reference in
\cref{fig:lin-emulate} shows the model $G\theta$. The red cross denotes the
corresponding observation. The solid lines correspond to the mean of the GP,
while the shaded regions contain $2$ standard deviations of predicted
variability. We can see in \cref{fig:lin-emulate} that the GP increases its
accuracy as the amount of training data is increased. In the end, for training
sets of size $J \geq 16$, it correctly simulates the linear model with low
uncertainty in the main support of the posterior.

\begin{figure}[!ht]
	\centering
	\includegraphics[width=.98\linewidth]{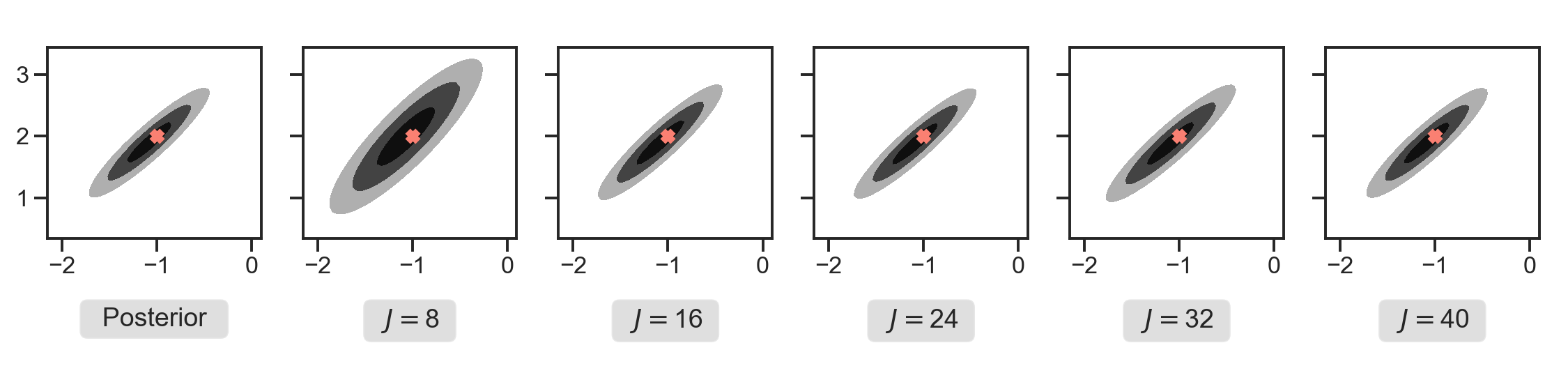}
	\caption{Density of a Gaussian with mean and covariance estimated from
	GP-based MCMC samples using $M=J$ design points. The true posterior distribution
	is shown in the far left. Each GP-based MCMC generated $2\times10^4$
	samples. These samples are not shown for clarity. }
	\label{fig:lin-sample}
\end{figure}

\cref{fig:lin-sample} depicts results in the sampling step \ceS. These are
obtained by using a GP approximation of $G\theta$ within MCMC. The GP-based MCMC
uses \eqref{eq:phi_gp2} since the forward model is deterministic and the data is
polluted by noise. The contour levels show a Gaussian distribution with mean and
covariance estimated from $N_s = 2 \times 10^4$ GP-based MCMC samples (not
shown) in each of the different ensemble settings. The results show that the
true posterior is captured with an ensemble size of $16$ or more.

\section{Darcy Flow} \label{sec:D}

In this section, we apply our methodology to a PDE nonlinear inverse problem
arising in porous medium flow: the determination of permeability from
measurements of the pressure.


\subsection{Elliptic Inverse Problem}
\label{ssec:TSD}

The forward problem is to find the pressure field $p$ in a porous medium defined
by the (for simplicity) scalar permeability field $a$. Given a scalar field $f$
defining sources and sinks of fluid, and assuming Dirichlet boundary conditions
on the domain $D=[0,1]^2$, we obtain the following elliptic PDE determining the
pressure from permeability:
\begin{subequations}
\begin{align}
-\nabla \cdot ( a(x) \nabla p(x)) &= f(x), \quad x \in D.\\
p(x) &=0, \qquad x \in \partial D.
\end{align}
\label{eq:darcy-system}
\end{subequations}
We assume that the permeability is dependent on unknown parameters $\theta \in
\R^p$, so that $a(x)=a(x;\theta).$ The inverse problem of interest is to
determine $\theta$ from noisy observations of $d$ linear functionals
(measurements) of $p(x;\theta)$. Thus,
\begin{equation}\label{eq:darcy-obs}
\G_{j}(\theta)=\ell_j\bigl(p(\cdot;\theta)\bigr) + \eta, \qquad j=1,\cdots, d.
\end{equation}
We assume the additive noise $\eta$ to be a mean zero Gaussian with
covariance equal to $\gamma^2 I.$
We also assume that $a \in L^{\infty}(D;\R)$ so that $p \in H^1_0(D;\R)$; thus,
the $\ell_j$ are linear functionals on the space $H^1_0(D;\R)$. In practice, we
work with pointwise measurements so that $\ell_j(p)=p(x_j)$. The implied linear
functionals are not elements of the dual space $H^1_0(D;\R)$ in dimension $2$
but mollifications of them are. In practice, mollification with a narrow kernel
does not affect results of the type presented here
\citep{iglesias2013evaluation}, and so we do not use it.

We introduce a log-normal parameterization of $a(x;\theta)$ as follows:
\begin{equation}
\label{eq:adefine}
\log a(x; \theta)=\sum_{\ell \in K_p} \theta_\ell \, \sqrt{\lambda_\ell}\, \varphi_{\ell}(x)
\end{equation}
where
\begin{align}
\label{eq:adefine0}
  \varphi_{\ell}(x)=\cos\bigl(\pi \langle \ell,x \rangle \bigr), \qquad
  \lambda_\ell = (\pi^2 |\ell|^2 + \tau^2)^{-\alpha},
\end{align}
and $K_p \subset K \equiv \Z^2$ is the set, with finite cardinality, of
indices over which the random series is summed. A priori we assume that $\theta_\ell \sim \N(0,1)$
so that we have Karhunen-Loeve representation of a Gaussian random field for $a$ \citep{pavliotis2014stochastic}. We often find it helpful to write
\eqref{eq:adefine} as a sum over a one-dimensional variable rather than a
lattice:
\begin{equation}
\label{eq:adefine2}
\log a(x; \theta')=\sum_{k \in Z_p} \theta_k' \, \sqrt{\lambda_k'}\, \varphi_{k}'(x).
\end{equation}
We order the indices in $Z_p \subset \mathbb{Z}^+$ so that the eigenvalues
$\lambda_k'$ are non-increasing with respect to $k$.

\subsection{Numerical Results}
\label{ssec:ND}

We generate an underlying true random field by sampling $\theta^\dagger \in
\R^p$ from a standard multivariate Gaussian distribution $\N(0, I_p)$, of
dimension $p = 2^8 = 256$. This is used as the coefficients in
\eqref{eq:adefine} by means of re-labelling \eqref{eq:adefine2}. We create data
$y$ from \eqref{eq:IP} with a random perturbation $\eta \sim \N(0,0.005^2 \times
I_d),$ where $I_d$ denotes the identity matrix. For the Bayesian inversion, we
consider using a truncation of \eqref{eq:adefine2} with $p'< p$ terms.
Specifically, we consider $p' = 10$ which will allow us to avoid the inverse
crime of using the same model that generated the data to solve the inverse
problem \citep{kaipio2006statistical}. We employ a non-informative centered
Gaussian prior with covariance $\Gammzero=10^2 \times I_{p'};$ this is also used
to initialize the ensemble for EKS. We consider ensembles of size $J \in \{128,
512\}$.

We perform the complete CES procedure starting with EKS as described above for
the calibration step \Ces. The emulation step \cEs\ uses a GP with a linear mean
Gaussian process with squared-exponential kernel \eqref{eq:gp_rbf}. Empirically,
the linear mean allows us to capture a significant fraction of the relevant
parameter response. The GP covariance matrix $\Gammgp(\theta)$ accounts for the
variability of the residuals from the linear function. The sampling step \ceS\
is performed using the Random Walk procedure described in \cref{ssec:sample},
where a Gaussian transition distribution is matched to the first two moments of
the ensemble at the last iteration of EKS. In this experiment, the likelihood
\eqref{eq:phi_gp2} is used because the forward model is a deterministic map, and
we have data polluted by additive noise.

\begin{figure}[!ht]
	\centering
	\subfigure[Original $\theta$.]{\includegraphics[height = .8\paperheight,width=.2\linewidth, keepaspectratio=true]{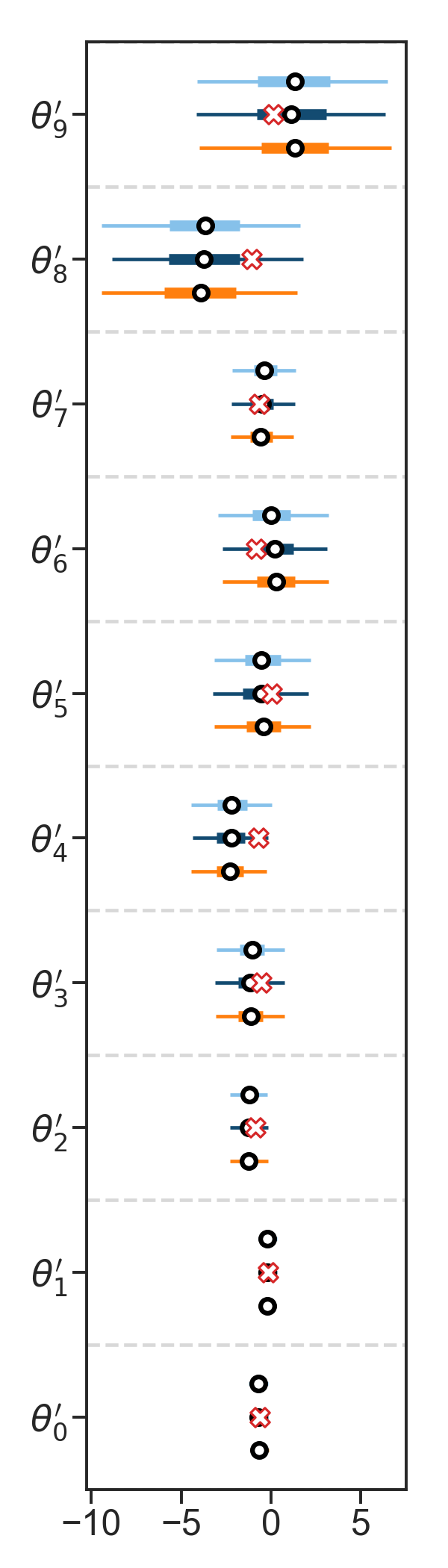}\label{sfig:darcy-fp}}\hfil
	\subfigure[Rescaled $\theta$.]{\includegraphics[height = .85\paperheight,width=.2\linewidth, keepaspectratio=true]{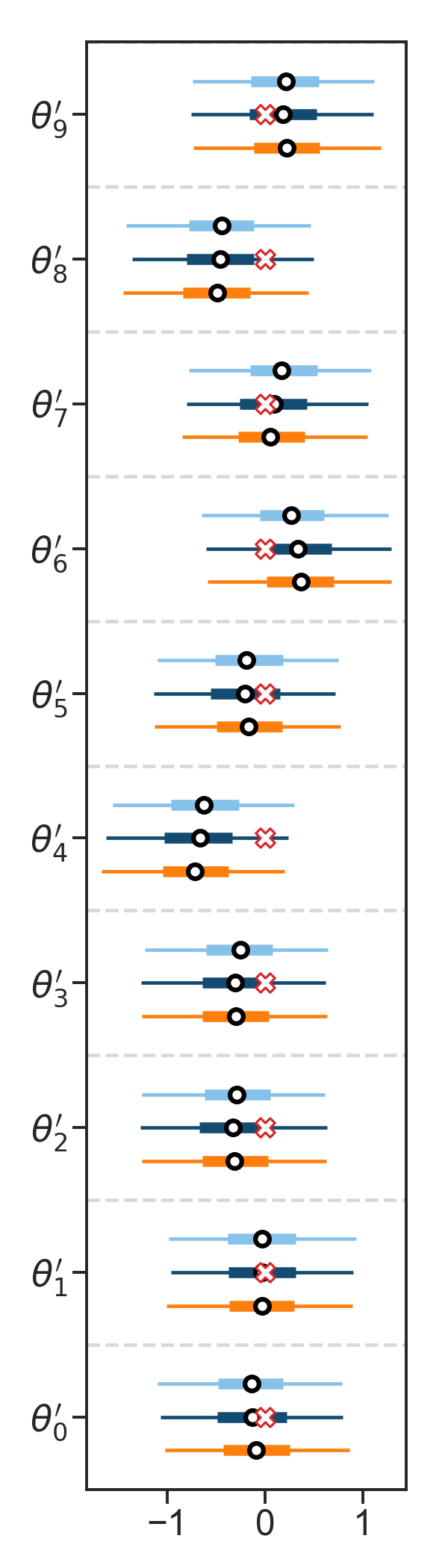}\label{sfig:darcy-fpstand}}\hfil
	\subfigure[MCMC traceplot]{\includegraphics[height = .39\paperheight,width=.52\linewidth]{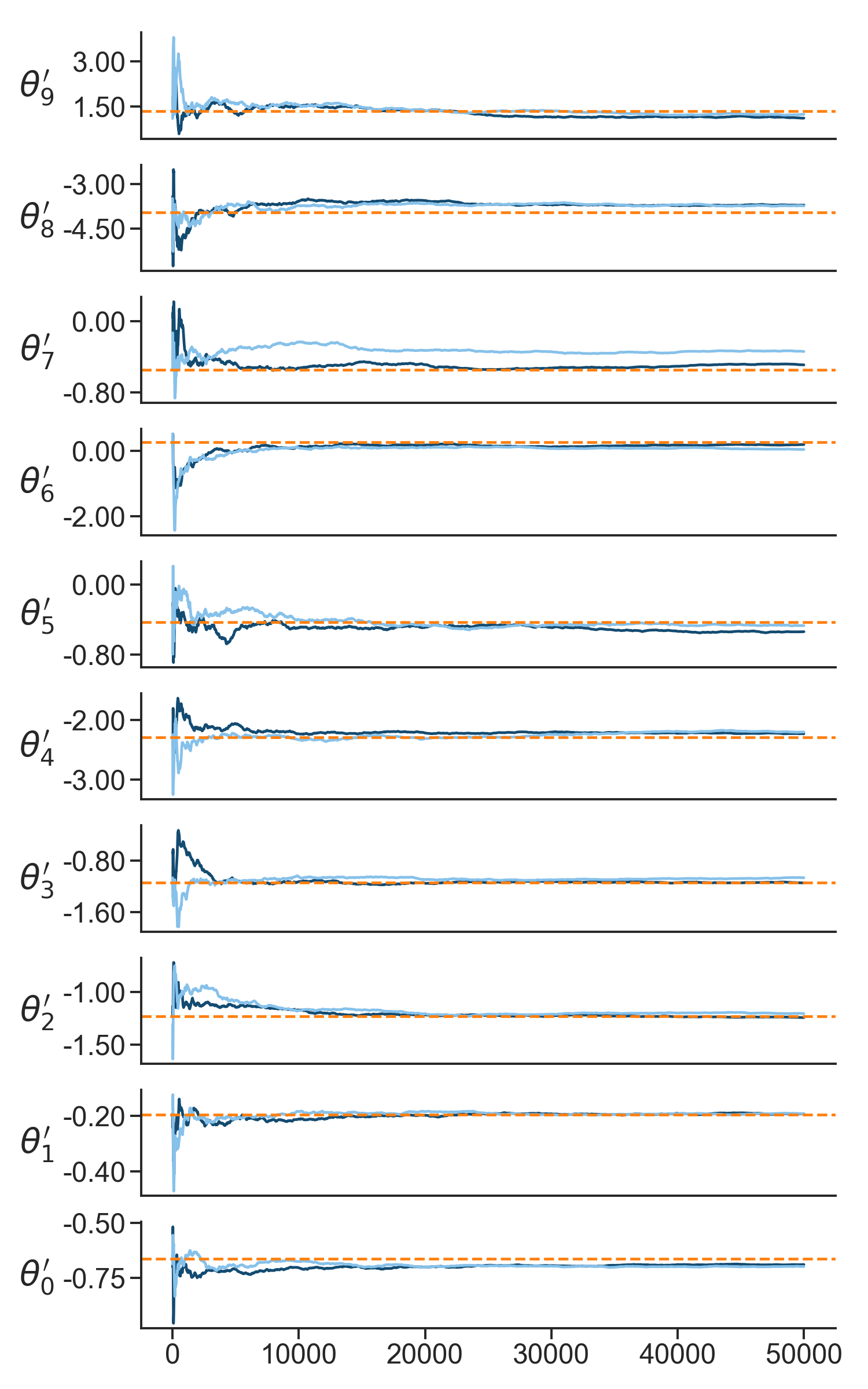}\label{sfig:darcy-rave}}\hfil
	\caption{Results of CES in the Darcy flow problem. Colors throughout the panel
	denote results using different calibration and GP training settings. This are:
	light blue -- ensemble of size $J = 128$; dark blue -- ensemble of size $J =
	512$; and orange, the MCMC gold standard. Left panel shows each $\theta$
	component for CES. The middle panel shows the same information, but using
	standardized components of 	$\theta$. The interquartile range is displayed
	with a thick line; the $95\%$ credible intervals with a thin line; and the
	median with circles. The right panel shows typical MCMC running averages,
	demonstrating stationarity of the Markov chain.
	}
 	\label{fig:darcy-ces}
\end{figure}

We compare the results of the CES procedure with those obtained from a gold
standard MCMC employing the true forward model. The results are summarized in
\cref{fig:darcy-ces}. The right panel shows typical MCMC running averages,
suggesting stationarity of the Markov chain. The left panel shows the forest
plot of each $\theta$ component. The middle panel shows the standardized
components of $\theta$. These forest plots show the interquartile range with a
thick line; the $95\%$ credible intervals with a thin line; and the median with
circles. The true value of the parameters are denoted by red crosses. The
results demonstrate that the CES methodology accurately reproduces the true
posterior using calibration and training with $M = J = 512$ ensemble members.
For the smaller ensemble, $ M = J = 128 $ there is a visible systematic
deviation in some components, like $\theta_7$. Although, the CES posterior does
capture the true value. This is in contrast to the gold standard MCMC, which
uses tens of thousands of evaluations.

\begin{figure}[!ht]
	\centering
	\subfigure[PDF - Exceedance on pressure field.]{\includegraphics[height = .15\paperheight, width=.45\linewidth]{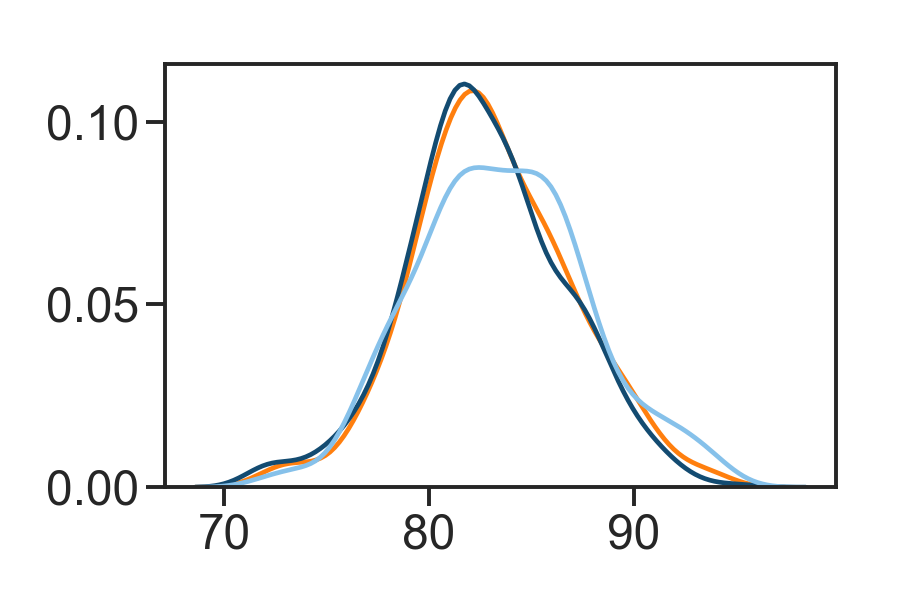}} \quad
	\subfigure[PDF - Exceedance on permeability field.]{\includegraphics[height = .15\paperheight, width=.45\linewidth]{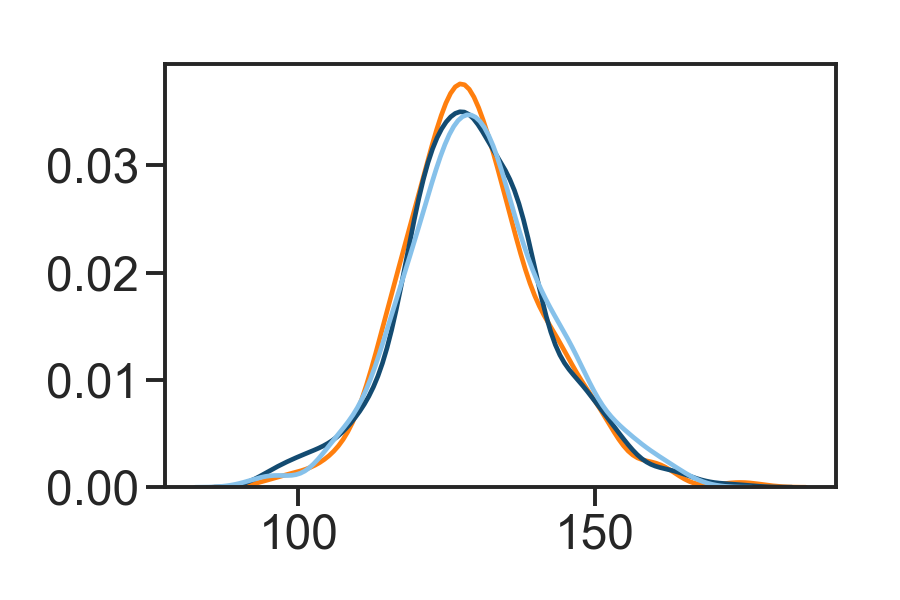}}
	\caption{Forward UQ of exceedance on both the pressure field, $p(\cdot) > \bar
	p$, and permeability field, $a(\cdot) > \bar a$. Both reference levels are
	obtained from the forward model at the truth $\theta^\dagger$ and taking the
	median across the locations \eqref{eq:darcy-obs}. The PDFs are constructed by
	running the forward model on a small set of samples, $N_{\mathsf{UQ}} = 250$,
	and computing the number lattice points that exceed the threshold. The samples
	are obtained by using the CES methodology (light blue -- CES with $ M = J =
	128$, and dark blue -- CES with $M = J = 512$). The samples in orange are
	obtained from a gold standard MCMC using the true forward model within the
	likelihood, rather than the emulator.}
	\label{fig:darcy-fwd-uq}
\end{figure}

The results from the CES procedure are also used in a forward UQ setting:
posterior variability in the permeability is pushed forward onto quantities of
interest. For this purpose, we consider exceedances of both the pressure and
permeability fields above certain thresholds. These thresholds are computed from
the observed data by taking the median across the $50$ available locations
\eqref{eq:darcy-obs}. The forward model \eqref{eq:darcy-system} is solved with
$N_{\mathsf{UQ}}=500$ different parameter settings coming from samples of the
CES Bayesian posterior. We also show the comparison with the gold standard
computed using the true forward model. The number of lattice points in the KL expansion
exceeding such threshold levels is computed and recorded for each sampled
parameter. \cref{fig:darcy-fwd-uq} shows the corresponding KDE for the
probability density function (PDF) in this forward UQ exercise. The orange lines
correspond to the PDF of the number of points in the lattice that exceed the
threshold computed from the samples drawn using MCMC with the Darcy flow model.
The corresponding PDFs associated to the CES posterior, based on calibration and
emulation using different ensemble sizes, are shown in different blue tonalities
(light blue -- CES with $ M = J = 128$, and dark blue -- CES with $M = J =
512$). A nonparametric $k$-sample Anderson--Darling test at $5\%$ significance
level for the $M=J=128$ case, shows evidence to reject the null hypothesis of
the samples being drawn from the same distribution in the pressure exceedance forward
UQ. In the other cases, such test does not provide statistical evidence to
reject the hypothesis that the distributions are similar to the one based on the
Darcy model itself.

\section{Time-Averaged Data}
\label{sec:T}

In parameter estimation problems for chaotic dynamical systems, such as those
arising in climate modeling
\citep{Emmet,jarvinen2012ensemble,schneider2017earth}, data may only be
available in time-averaged form; or it may be desirable to study time-averaged
quantities in order to ameliorate difficulties arising from the complex
objective functions, with multiple local minima, which arise from trying to
match trajectories \citep{abarbanel2013predicting}. Indeed the idea fits the more
general framework of feature-based data assimilation introduced in
\citep{morzfeld2018feature} which, in turn, is closely related to the idea of
extracting sufficient statistics from the raw data
\citep{fisher1922mathematical}. The methodology developed in this section
underpins similar work conducted for a complex climate model described in the
paper \citep{Emmet}.


\subsection{Inverse Problems From Time-Averaged Data}
\label{ssec:TS}

The problem is to estimate the parameters $\theta \in \R^p$ of a dynamical
system evolving in $\R^m$ from data $y$ comprising time-averages of an
$\R^d-$valued function $\varphi(\cdot).$ We write the dynamical system as
\begin{equation} \dot{z} = F(z; \theta), \quad z(0) = z_0. \label{eq:ode}
\end{equation} Since $z(t)\in \R^m$ and $\theta \in \R^p$ we have $F: \R^m
\times \R^p \to \R^m$ and $\varphi: \R^m \to \R^d.$ We will write $z(t;\theta)$
when we need to emphasize the dependence of a trajectory on $\theta.$ In view of
the time-averaged nature of the data it is useful to define the operator
\begin{equation}\label{eq:time-average}
	\mathcal{G}_\tau(\theta; z_0) = \frac{1}{\tau}\int_{T_0}^{T_0 + \tau}\varphi(z(t;\theta))dt
\end{equation}
where $T_0$ is a predetermined spinup time, $\tau$ is the time horizon over
which the time-averaging is performed, and $z_0$ the initial condition of the
trajectory used to compute the time-average. Our approach proceeds under the
following assumptions:
\nc
\begin{assumption}

The dynamical system \eqref{eq:ode} satisfies:

\begin{enumerate}

		\item For every $\theta \in \Theta,$ \eqref{eq:ode} has a compact attractor
		$\mathcal{A}$, supporting an invariant measure $\mu(dz;\theta)$. The system
		is ergodic, and the following limit -- a Law of Large Numbers (LLN) analogue
		-- is satisfied: for $z_0$ chosen at random according to measure
		$\mu(\cdot;\theta)$ we have, with probability one,
		\begin{equation}
		\lim_{\tau\rightarrow\infty} \mathcal{G}_\tau(\theta; z_0) = \mathcal{G}(\theta):=\int_{\mathcal{A}} \varphi(z)\mu(dz;\theta).
		\end{equation}
		\item We have a Central Limit Theorem (CLT) quantifying the ergodicity: for
		$z_0$ distributed according to $\mu(dz;\theta),$
		\begin{align}
				\G_\tau({\theta};z_0)& \approx \G({\theta})+\frac{1}{\sqrt{\tau}}\N(0,\Sigma(\theta)). \label{eq:clt}
		\end{align}
\end{enumerate}
\end{assumption}
In particular, the initial condition plays no role in time averages over the
infinite time horizon. However, when finite time averages are employed,
different random initial conditions from the attractor give different random
errors from the infinite time-average, and these, for fixed spinup time $T_0$,
are approximately Gaussian. Furthermore, the covariance of the Gaussian depends
on the parameter $\theta$ at which the experiment is conducted. This is
reflected in the noise term in \eqref{eq:clt}.

Here we will assume that the model is perfect in the sense that the data we are
presented with could, in principle, be generated by \eqref{eq:ode} for some
value(s) of $\theta$ and $z_0.$ The only sources of uncertainty come from the
fact that the true value of $\theta$ is unknown, as is the initial condition
$z_0.$ In many applications, the values of $\tau$ that are feasible are limited
by computational cost. The explicit dependence of $\G_\tau$ on $\tau$ serves to
highlight the effect of $\tau$ on the computational budget required for each
forward model evaluation. Use of finite time-averages also introduces the
unwanted nuisance variable $z_0$ whose value is typically unknown, but not of
intrinsic interest. Thus, the inverse problem that we wish to solve is to find
$\theta$ solving the equation
\begin{equation}
\label{eq:yT}
y = \G_T(\theta; z_0)
\end{equation}
where $z_0$ is a latent variable and $T$ is the computationally-feasible time
window we can integrate the system \eqref{eq:ode}. We observe that the preceding
considerations indicate that it is reasonable to assume that
\begin{equation}
y = \mathcal{G}(\theta) + \, \eta \,, \label{eq:yT2}
\end{equation}
where $\eta \sim \N(0,\Gammy(\theta))$ and $\Gammy(\theta)=T^{-1}\Sigma(\theta).$
We will estimate $\Gammy(\theta)$ in two ways: firstly using long-time series
data; and secondly using a GP informed by forward model evaluations.

We first estimate $\Gammy(\theta)$ directly from $\mathcal{G}_\tau$ with $\tau
\gg T.$ We will not employ $\theta-$dependence in this setting and simply
estimate a fixed covariance $\Gammo.$ This is because, in the applications we
envisage such as climate modeling \citep{Emmet}, long time-series data over
time-horizon $\tau$ will typically be available only from observational data.
The cost of repeatedly simulating at different candidate $\theta$ values is
computationally prohibitive, in contrast, to simulations over a shorter
time-horizon $T$. We apply EKS to make an initial calibration of $\theta$ from
$y$ given by \eqref{eq:yT}, using $\G_T(\theta^{(j)};z_0^{(j)})$ in place of
$\G(\theta^{(j)})$ and $\Gammo$ in place of $\Gammy(\cdot)$, within the
discretization \eqref{eq:implicit} of \eqref{eq:implement}. We find the method
to be insensitive to the exact choice of $z_0^{(j)}$, and typically use the
final value of the dynamical system computed in the preceding step of the
ensemble Kalman iteration. We then take the evaluations of $\mathcal{G}_T$ as
noisy evaluations of $\mathcal{G}$, from which we learn the Gaussian process
$\mathcal{G}^{(M)}$. We use the mean $m(\theta)$ of this Gaussian process as an
estimate of $\mathcal{G}.$ Our second estimate of $\Gammy(\theta)$ is obtained
by using the covariance of the Gaussian process $\Gammgp(\theta)$. We can
evaluate the misfit through either of the expressions
\begin{subequations}
\label{eq:GPMF}
\begin{align}
	\Phim (\theta;y) & = \frac12\|y - m(\theta)\|^2_{\Gammo}, \label{eq:ta-phim}\\
	\Phigp(\theta;y) & = \frac12\|y - m(\theta)\|^2_{\Gammgp(\theta)} + \frac12 \log \det\Gammgp(\theta). \label{eq:ta-phigp}
\end{align}
\end{subequations}
Note that equations \eqref{eq:GPMF} are the counterparts of
\eqref{eq:phi_gp} and \eqref{eq:phi_m} in the setting with time-averaged data.
In what follows, we will contrast these misfits, both based on the learnt GP
emulator, with the misfit that uses the noisy evaluations $\mathcal{G}_T$
directly. That is, we use the misfit computed as
\begin{equation}
\label{eq:OMF}
	\Phi_T(\theta;y) = \frac12\|y - \mathcal{G}_T(\theta)\|^2_{\Gammo}.
\end{equation}
In the latter, dependence of $\G_T$ on initial conditions is suppressed.

\subsection{Numerical Results -- Lorenz '63}
\label{ssec:L63}

We consider the 3-dimensional Lorenz equations \citep{lorenz1963}
\begin{subequations}
\label{eq:lor63}
\begin{align}
	\dot x_1 &= \sigma (x_2 - x_1), \\
	\dot x_2 &= r x_1 - x_2 - x_1 x_3, \\
	\dot x_3 &= x_1x_2 - bx_3,
\end{align}
\end{subequations}
with parameters $\sigma, b, r \in \R_+$. Our data is found by simulating
\eqref{eq:lor63} with $(\sigma, b, r ) = (10, 28, 8/3)$, a value at which the
system exhibits chaotic behavior. We focus on the inverse problem of recovering
$(r, b)$, with $\sigma$ fixed at its true value of $10$, from
time-averaged data.

\begin{figure}[!ht]
	\includegraphics[width=.80\linewidth]{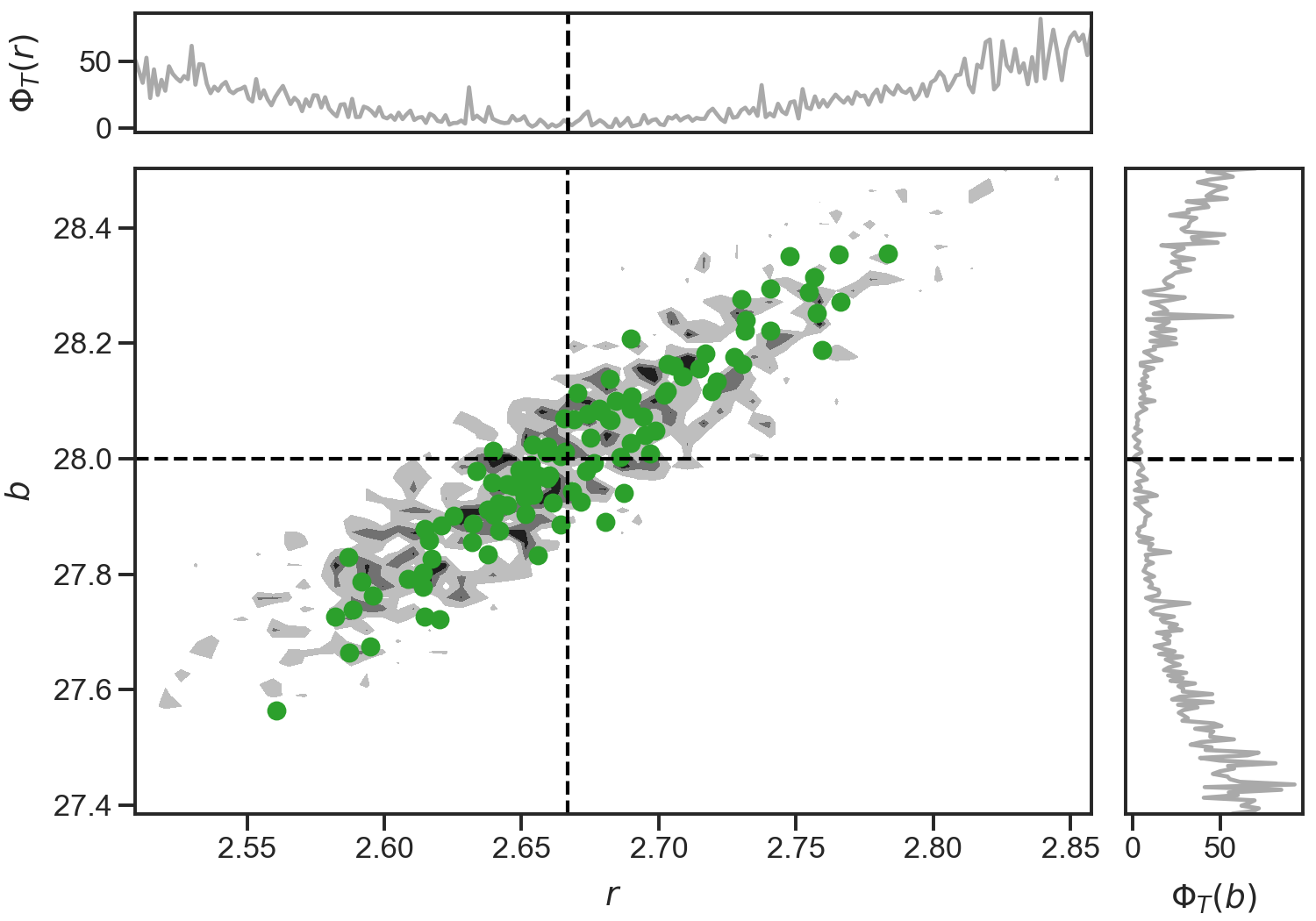}
	\caption{Contour levels of the misfit of the Lorenz '63 forward model
	corresponding to $(67\%, 90\%, 99\%)$ density levels. The dotted lines shows
	the locations of the true parameter values that generated the data. The green
	dots shows the final ensemble of the EKS algorithm. The marginal plots show
	the misfit as a 1-d function keeping one parameter fixed at the truth while varying the
	other. This highlights the noisy response from the time-average forward
	model $\G_T$.}
	\label{fig:l63-misfit}
\end{figure}

Our statistical observations are first and second moments over time windows of
size $T=10$. Our vector of observations is computed by taking $\varphi:\R^3
\mapsto \R^9$ to be
\begin{align}
\varphi(x)=(x_1,x_2,x_3,x_1^2,x_2^2,x_3^2,x_1x_2,x_2x_3,x_3x_1).
\end{align}
This defines $\mathcal{G}_T$. To compute $\Gammo$ we used
time-averages of $\varphi(x)$ over $\tau = 360$ units of time, at the true value
of $\theta$; we split the time-series into windows of size $T$ and neglect an
initial spinup of $T_0 = 30$ units of time. Together $\mathcal{G}_T$ and $\Gammo$
produce a noisy function $\Phit$ as
depicted in \cref{fig:l63-misfit}. The noisy nature of the energy
landscape, demonstrated in this figure, suggests that standard optimization and
MCMC methods may have difficulties; the use of GP emulation will act to smooth
out the noise and lead to tractable optimization and MCMC tasks.


For the calibration step \Ces, we run the EKS using the estimate of
$\Gamma=\Gammo$ within the algorithm \eqref{eq:implement}, and within the misfit
function \eqref{eq:OMF}, as described in \cref{ssec:TS}. We assumed the
parameters to be {\em a priori} governed by an isotropic Gaussian prior in
logarithmic scale. The mean of the prior is $\mupost_0 = (3.3, 1.2)^\top$ and
its covariance is $ \Sigma_0 = \rm{diag} (0.15^2, 0.5^2)$. This gives broad
priors for the parameters with $99\%$ probability mass in the region $[20, 40]
\times [0, 15].$ The results of evolving the EKS through $11$ iterations can be
seen in \cref{fig:l63-misfit}, where the green dots represent the final
ensemble. The dotted lines locate the true underlying parameters in the $(r,b)$
space.

\begin{figure}[!ht]
	\centering
	\includegraphics[width=.9\linewidth]{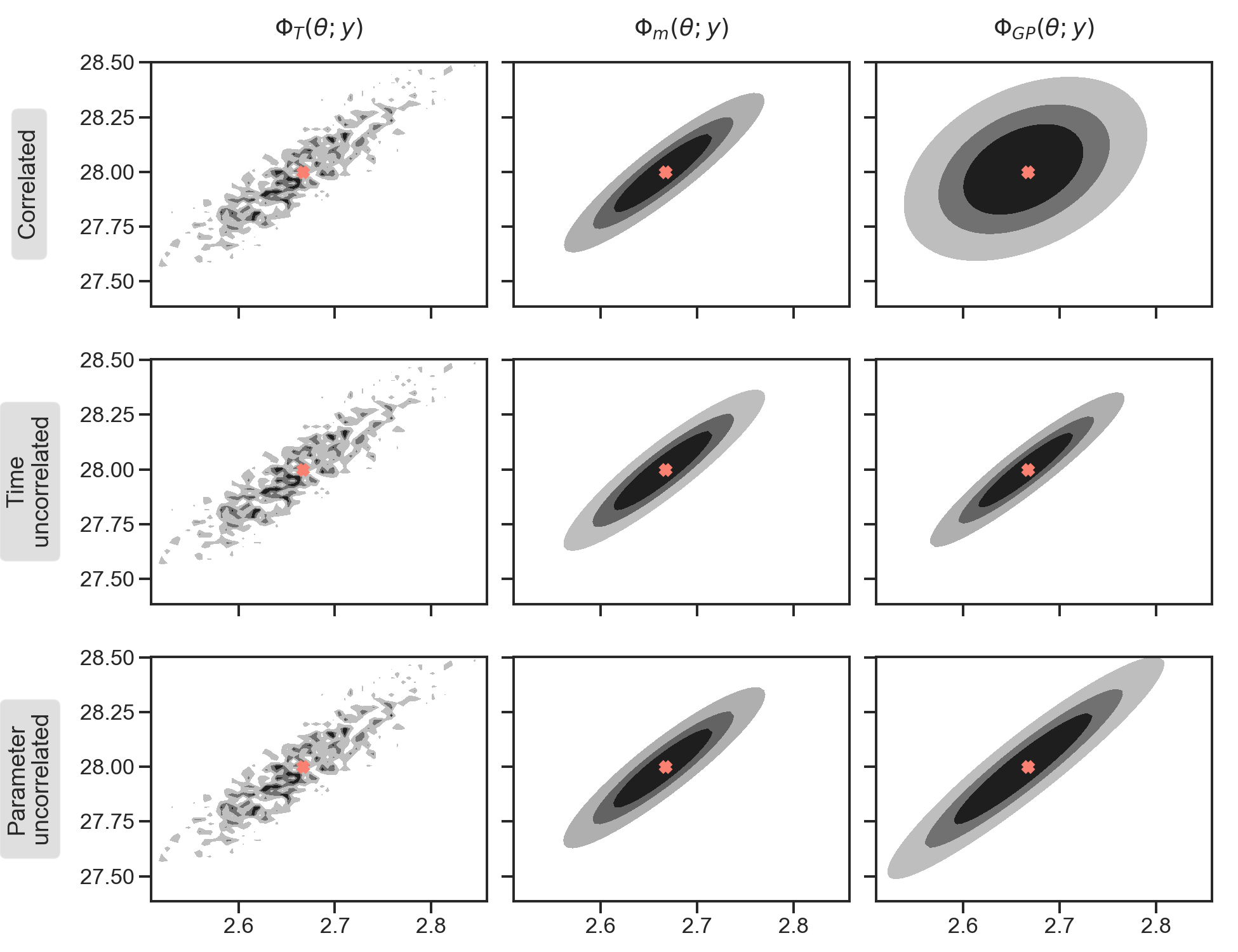}
	\caption{Contour levels of the Lorenz '63 posterior distribution corresponding
	to $(67\%, 90\%, 99\%)$ density levels. For each row we depict: in the left
	panel, the contours using the true forward model; in the middle panel, the
	contours of the misfit computed as $\Phim$ \eqref{eq:ta-phim}; and in the
	right panel, the contours of the misfit obtained using $\Phigp$
	\eqref{eq:ta-phigp}. The difference between rows is due to the decorrelation
	strategy used to learn the GP emulator, as indicated in the leftmost labels.
	The GP-based densities show an improved estimation of uncertainty arising from GP
	estimation of the infinite time-averages, in comparison
	with employing the noisy exact finite time averages, for both
	decorrelation strategies.}
	\label{fig:l63-emulator}
\end{figure}

For the emulation step \cEs, we use GP priors for each of the $9$ components of
the forward model. The hyper-parameters of these GPs are estimated using
empirical Bayes methodology. The $9$ components do not interact and are treated
independently. We use only the input-output pairs obtained from the last
iteration of EKS in this emulation phase, although earlier iterations could also
have been used. This choice focuses the training runs in regions of high
posterior probability. Overall, the GP allows us to capture the underlying
smooth trend of the misfit. In \cref{fig:l63-emulator} (top row) we show (left
to right) $\Phi_T, \Phim$, and $\Phigp$ given by
\eqref{eq:GPMF}--\eqref{eq:OMF}. Note that $\Phim$ produces a smoothed version
of $\Phi_T$, but that $\Phigp$ fails to do so -- it is smooth, but the
orientations and eccentricities of the contours are not correctly captured. This
is a consequence of having only diagonal information to replace the full
covariance matrix $\Gamma$ by $\Gamma(\theta)$ and not learning dependencies
between the $9$ simulator outputs that comprise $\mathcal{G}_T.$

We explore two options to incorporate output dependency. These are detailed in
\cref{apx:diag}, and are based on changing variables according to either a
diagonalization of $\Gammo$ or on an SVD of the centered data matrix formed from
the EKS output used as training data $\{\G_T(\theta^{(i)})\}_{i=1}^{M}$. The
effect on the emulated misfit when using these changes of variables is depicted
in the middle and bottom rows of \cref{fig:l63-emulator}. We can see that the
misfit $\Phim$ \eqref{eq:ta-phim} respects the correlation structure of the
posterior. There is no notable difference between using a GP emulator in the
original or decorrelated output system. This can been seen in the middle column
in \cref{fig:l63-emulator}. However, if the variance information of the
GP emulator is introduced to compute $\Phigp$ \eqref{eq:ta-phigp}, decorrelation
strategies allows us to overcome the problems caused when using diagonal
emulation.

\begin{figure}[!ht]
	\includegraphics[width=.9\linewidth]{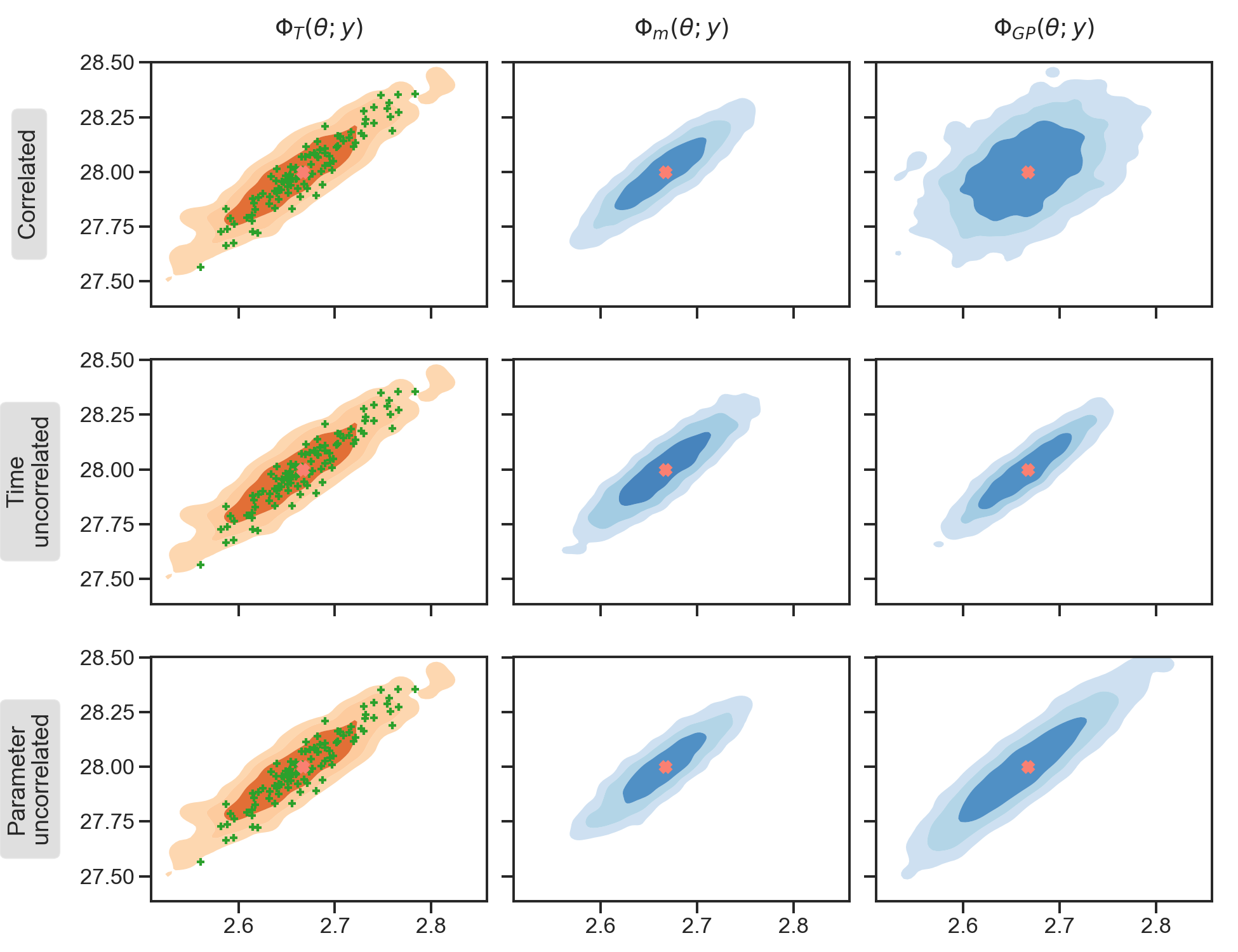}
	\caption{Samples using different modalities of the GP emulator. The orange
	kernel density estimate (KDE) is based on samples from random walk Metropolis
	using the true forward model. The blue contour levels are KDE using the
	GP-based MCMC. All MCMC-based KDE approximate posteriors are computed from
	$N_s = 20,000$ MCMC samples. The green dots in the left panels depict the final
	ensemble from the EKS as a comparison. Furthermore, the CES-based densities
	are computed more easily as the MCMC samples decorrelate more rapidly due to a
	higher acceptance probability for the same size of proposed move.}
	\label{fig:l63-mcmc}
\end{figure}

Finally, the sample step \ceS\ is performed using the GP emulator to accelerate
the sampling and to correct for the mismatch of the EKS in approximating the
posterior distribution, as discussed in \cref{ssec:sample}. In this section,
random walk metropolis is run using 5,000 samples for each setting -- using the
misfits $\Phi_T$, $\Phim$ or $\Phigp$. The Markov chains are initialized at the
mean of the last iteration of the EKS. The proposal distribution used for the
random walk is a Gaussian with covariance equal to the covariance of the
ensemble at said last iteration. The samples are depicted in
\cref{fig:l63-mcmc}. The orange contour levels represent the kernel density
estimate (KDE) of samples from a random walk Metropolis algorithm using the true
forward model. On the other hand the blue contour levels represent the KDE of
samples using $\Phim$ or $\Phigp$, equations \eqref{eq:ta-phim} or
\eqref{eq:ta-phigp} respectively. The green dots in the left panels depict the
final ensemble from EKS. It should be noted that using $\Phim$ for MCMC has an
acceptance probability of around $41\%$ in each of the emulation strategy
(middle column). The acceptance rate increases slightly to around $47\%$ by
using $\Phigp$ (right column). The original acceptance rate is $16\%$ if the
true forward model is employed. The main reason is the noisy landscape of the
posterior distribution. In this experiment, the use of a GP emulator showcases
the benefits of our approach as it allows to generate samples from the posterior
distribution more efficiently than standard MCMC, not only because the emulator
is faster to evaluate, but also because it smoothes the log-likelihood. Careful
attention to how the emulator model is constructed and the use of nearly
independent co-ordinates in data space, helps to make the approximate
methodology viable.

\subsection{Numerical Results -- Lorenz '96}
\label{ssec:L96}

We consider the multiscale Lorenz '96 model \citep{lorenz1996}. This model
possesses properties typically present in the earth system
\citep{schneider2017earth} such as advective energy conserving nonlinearity,
linear damping and large scale forcing, and multiscale coexistence of slow and
fast variables. It comprises $K$ slow variables $X_k$ $(k = 1, \ldots K)$, each
coupled to $L$ fast variables $Y_{l,k}$ $(l = 1,\ldots,L)$. The dynamical system
is written as
\begin{subequations}\label{eq:lorenz96}
	\begin{align}
		\frac{\rd X_k}{\rd t} & = - X_{k-1} \left(X_{k-2} - X_{k+1} \right) - X_k + F - h c \, \bar Y_k \label{eq:l96-slow}\\
		\frac 1 c \frac{\rd Y_{l,k}}{\rd t} & = -b Y_{l+1,k}\left(Y_{l+2,k} - Y_{l-1,k}\right) - Y_{l,k} + \frac h L \, X_k, \label{eq:l96-fast}
	\end{align}
\end{subequations}
where $\bar Y_k = \frac1L \sum_{ l= 1}^L Y_{l,k}$. The slow and fast variables
are periodic over $k$ and $l$, respectively. This means that $X_{k+K} = X_k,$
$Y_{l,k+K} = Y_{l,k},$ and $Y_{l+L,k} = Y_{l,k+1}.$ A geophysical interpretation
of the model may be found in \citep{lorenz1996}.

The scale separation parameter, $c,$ is naturally constrained to be a
non-negative number. Thus, our methods consider the vector of parameters $\theta
\coloneqq (h, F, \log c, b)$. We perform Bayesian inversion for $\theta$ based
on data averaged across the $K$ locations and over time windows of length $T =
100.$ To this end, we define our $k-$indexed observation operator $\varphi_k:
\R \times \R^{L} \mapsto \R^{5}$, by
\begin{align}
	\varphi_k(Z) \coloneqq \varphi(X_k, Y_{1,k}, \ldots, Y_{L,k}) = \left( X_k, \bar Y_k,
	X_k^2, X_k \bar Y_k, \bar Y^2_k \right),
\end{align}
where $Z$ denotes the state of the system (both fast and slow variables) for $k
= 1, \ldots, K.$ Then we define the forward operator to be
\begin{align}
	\G_T(\theta) = \frac1T \int_0^T \left( \frac1K \sum_{k = 1}^K \varphi_k(Z(s)) \right) ds.
\end{align}
With this definition, the data we consider is denoted by $y$ and uses the true
parameter $\theta^\dagger = (1, 10, \log 10, 10).$ As in the previous
experiment, a long simulation of length $\tau = \mathcal{O}(4 \times 10^4)$ is
used to compute the empirical covariance matrix $\Gammo$. This simulation window
$(\tau)$ is long enough to reach statistical equilibrium. The covariance
structure enables quantification of the finite time fluctuations around the
long-term mean. In the notation of \cref{ssec:TS}, we have the inverse problem
of using data $y$ of the form
\begin{align}
	y = \G_{T}(\theta^\dagger) + \eta,
\end{align}
where $T$ is the finite time-window horizon, and the noise is approximately
$\eta \sim \N(0, \Gammy)$. The prior distribution used for Bayesian inversion
assumes independent components of $\theta$. More explicitly, it assumes a
Gaussian prior with mean $\mupost_\theta = (0, 10, 2, 5)^\top$ and covariance
$\Gammzero = \text{diag}(1, 10, .1, 10)$.

\begin{figure}[!ht]
	\centering
	\includegraphics[width=.99\linewidth]{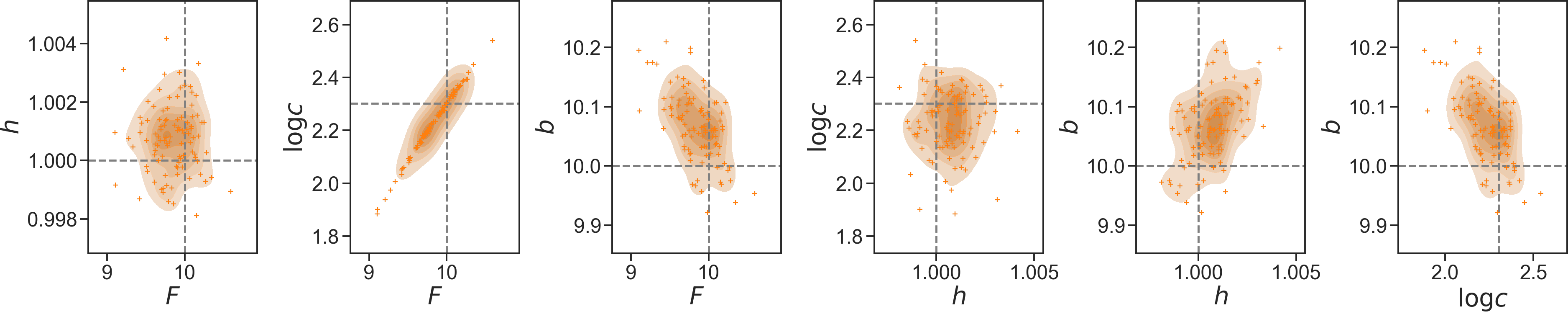}
	\caption{Samples and kernel density estimates of EKS applied to the Lorenz '96
	inverse problem. The ensemble, $J = 100$, shown corresponds to the last
	iteration. }
	\label{fig:l96-calibrate}
\end{figure}

The calibration step \Ces\ is performed using EKS as described in
\cref{ssec:calibrate}. The EKS algorithm is run for $54$ iterations with an
ensemble of size $J = 100$ which is initialized by sampling from the prior
distribution. The results of the calibration step are shown in
\cref{fig:l96-calibrate} as both bi-variate scatter plots and kernel density
estimates of the ensemble at the last iteration.

\begin{figure}[!ht]
	\includegraphics[width=0.99\linewidth]{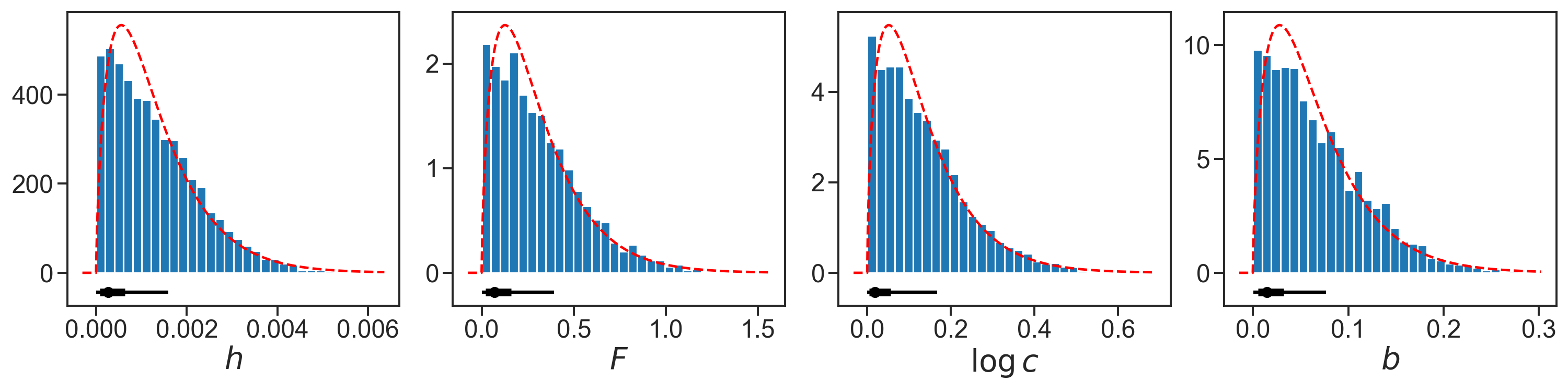}
	\caption{Histograms of pairwise distances for every component of the unknown
	parameters $\theta$ using the last iteration of EKS. Red dashed lines show the
	kernel density estimate of the histograms. The black box plot at the bottom
	shows the elicited GP--lengthscale priors. These priors are chosen to allow
	the GP kernel to decay rapidly from the training data; and to avoid the
	prediction of spurious short-term variations.}
	\label{fig:l96-emulate}
\end{figure}

The emulation step \cEs\ uses a subset of the trajectory of the ensemble as a
training set to learn a GP emulator. The trajectory is sampled in time under the
dynamics \eqref{eq:implement}, in such a way that we gather 10 different
snapshots of the ensemble. This is done by saving the ensemble every $6$
iterations of EKS. This gives $M = 10^3$ training points for the GP. Note that
\cref{fig:l96-calibrate} shows that each of the individual components of
$\theta$ has a different scale. We use a Gamma distribution as a prior for each
of the lengthscales to inform the GP of realistic sensitivities of the
space--time averages with respect to changes in the parameters. We use the last
iteration of EKS to inform such priors, as it is expected that the posterior
distribution will exhibit similar behaviour. The GP--lengthscale priors are
informed by the pairwise distances among the ensemble members, shown as
histograms in \cref{fig:l96-emulate}. The red dashed lines show the kernel
density estimates of such histograms. The black boxplots in the x-axes in
\cref{fig:l96-emulate} show the elicited priors found by matching a Gamma
distribution with $95\%$ percentiles equal to both a tenth of the minimum
pairwise distances, and a third of the maximum pairwise distances in each
component. These are chosen to allow the GP kernel to decay away from the
training data; and to avoid the prediction of spurious short-term variations.

\begin{figure}[!ht]
	\includegraphics[width=\linewidth]{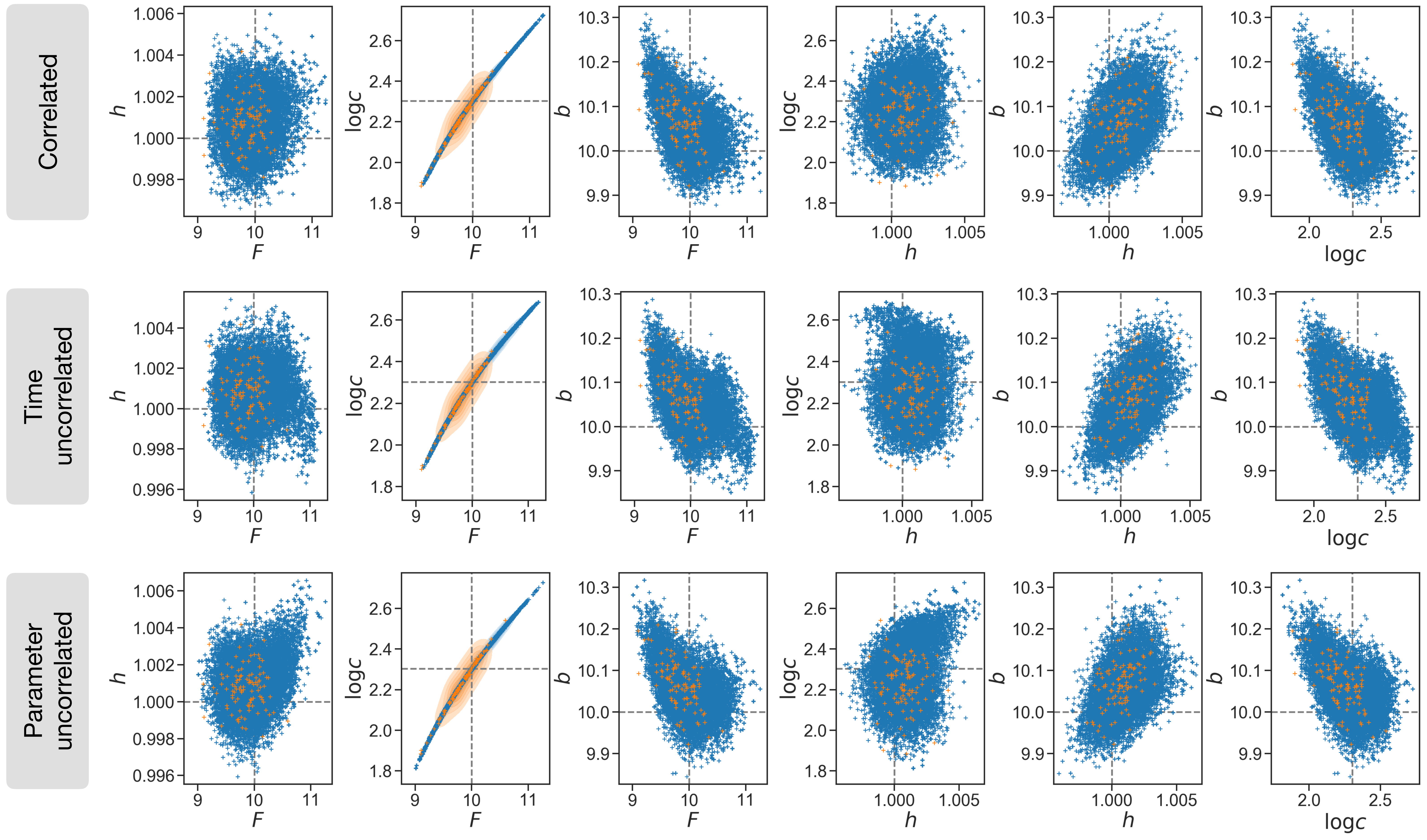}
	\caption{Shown in blue are the bi-variate scatter plots of the GP-based random
	walk metropolis using $N_s = 10^5$ samples. The orange dots are used as a
	reference and they correspond to the EKS' last iteration from the calibration
	step \Ces\ using an ensemble of size $J=100.$}
	\label{fig:l96-sample}
\end{figure}

As in the Lorenz '63 setting, we tried different emulation strategies for the
multioutput forward model. Independent GP models are fitted to the original
output and to the decorrelated output components based on both the
diagonalization of $\Gammo$ and SVD applied to the training data points, as
outlined in \cref{apx:diag}. The results shown in \cref{fig:l96-sample} are
achieved with zero mean GPs in both the original and time-diagonalized outputs.
For the SVD decorrelation, a linear mean GP was able to produce better
bi-variate scatter plots of $\theta$ in the sample step \ceS. That is, the
resulting bi-variate scatter plots of $\theta$ resembled better the last
iteration of EKS -- understood as our best guess of the posterior distribution.
For all GP settings, an identifiable Mat{\'e}rn kernel was used with smoothness
parameter $5/2$.

The sample step \ceS\ uses the GP emulator trained in the step above. We have
found in this experiment that using $\Phim$ for the likelihood term gave the
closest scatter plots to the EKS output. We did not make extensive studies with
$\Phigp$ as we found empirically that the additional uncertainty incorporated in
the GP-based MCMC produces an overly dispersed posterior, in comparison with EKS
samples, for this numerical experiment. The bi-variate scatter plots of $\theta$
shown in \cref{fig:l96-sample} show $N_s = 10^5$ samples using random walk
Metropolis with a Gaussian proposal distribution matched to the moments of the
ensemble at the last iteration of EKS. It should be noted that for this
experiment we could not compute a gold standard MCMC as we did in the previous
section. This is because of the high rejection rates and increased computational
cost associated with running a typical MCMC algorithm using the true forward
model. These experiments with Lorenz '96 confirm the viability of the CES
strategy proposed in this paper in situations where use of the forward model is
prohibitively expensive.

\section{Conclusions}
\label{sec:C}

In this paper, we have proposed a general framework for Bayesian inversion in
the presence of expensive forward models where no derivative information is
available. Furthermore, the methodology is robust to the possibility that only
noisy evaluations of the forward model are available. The proposed CES
methodology comprises three steps: calibration (using ensemble
Kalman---EK---methods), emulation (using Gaussian processes---GP), and sampling
(using Markov chain Monte Carlo---MCMC). Different methods can be used within
each block, but the primary contribution of this paper arises from the fact that
the ensemble Kalman sampler (EKS), used in the calibration phase, both locates
good parameter estimates from the data and provides the basis of an experimental
design for the GP regression step. This experimental design is well-adapted to
the specific task of Bayesian inference via MCMC for the parameters. EKS
achieves this with a small number of forward model evaluations, even for
high-dimensional parameter spaces, which accounts for the computational
efficiency of the method.

There are many future directions stemming from this work:
\begin{itemize}
	\item Combine all three pieces of CES as a single algorithm by interleaving
	the emulation step within the EKS, as done in iterative emulation techniques
	such as history matching.
	\item Develop a theory that quantifies the benefits of experimental design,
	for the purposes of Bayesian inference, based on samples that concentrate
	close to where the true Bayesian posterior concentrates.
	\item GP emulators are known to work well with low-dimensional inputs, but
	less well for the high-dimensional parameter spaces that are relevant in some
	application domains. Alternatives include the use of neural networks, or
	manifold learning to represent lower-dimensional structure within the input
	parameters and combination with GP.
	\item Deploying the methodology in different domains where large-scale
	expensive legacy forward models need to be calibrated to data.
\end{itemize}

{\footnotesize
\noindent
\textbf{Acknowledgements:}
 All authors are supported by the generosity of Eric and Wendy Schmidt by recommendation of the Schmidt Futures program, by Earthrise Alliance, Mountain Philanthropies, the Paul G. Allen Family Foundation, and the National Science Foundation (NSF, award AGS‐1835860). A.M.S. is also supported by NSF (award DMS-1818977) and by the Office of Naval Research (award N00014-17-1-2079)
}

\bibliography{main}
\bibliographystyle{abbrvnat}

\appendix

\section{Schemes To Find Uncorrelated Variables}\label{apx:diag}

\subsection{Time variability decorrelation}\label{apx:time-diag}

We present here a strategy to decorrelate the outputs of the forward model. It
is based on the noise structure of the available data. Here we assume that we
have access to $\Gammo$ and that it is diagonalized in the form
\begin{align} \Gammo = Q
\, \tilde \Gamma_{\scriptscriptstyle \textsf{obs}} \, Q^\top
\end{align}
The matrix $ Q\in \mathbb{R}^{d \times d}$ is orthogonal, and $\tilde \Gamma \in
\mathbb{R}^{d \times d}$ is an invertible diagonal matrix. Recalling that
$y=\mathcal{G}(\theta) + \eta$, and defining both $\tilde y = Q^\top y$ and
$\tilde \G(\theta) = Q^\top \G(\theta)$, we emulate the components of $\tilde
\G(\theta)$ as uncorrelated GPs. Recall that we are given $M$ training pairs
$\{\theta^{(i)},\G(\theta^{(i)})\}_{i=1}^{M}$. We transform these to data of the
form $\{\theta^{(i)},Q^\top\G(\theta^{(i)})\}_{i=1}^{M}$, which we emulate to
obtain
\begin{align}
	\tilde \G (\theta) \sim \N \left(\tilde m (\theta), \tilde \Gamma (\theta)\right).
\end{align}
This can be transformed back to the original output coordinates as
\begin{align}
	\G (\theta) \sim \N \left( Q \, \tilde m (\theta), Q \, \tilde \Gamma (\theta) \, Q^\top \right).
\end{align}
Using the resulting emulator, we can compute the misfit \eqref{eq:phi_gp} as
follows:
\begin{align}
\Phigp(\theta;y) = \frac12 \|\tilde y - \tilde m(\theta) \|^2_{\tilde \Gamma(\theta)} + \frac12 \log\det \tilde \Gamma(\theta).
\end{align}
Analogous considerations can be used to evaluate \eqref{eq:phi_m} or
\eqref{eq:phi_gp2}.

\subsection{Parameter variability decorrelation}\label{apx:pca-diag}

An alternative strategy to decorrelate the outputs of the forward model is
presented. It is based on evaluations of the simulator rather than the noise
structure of the data. As before, let us denote the set of $M$ available
input-output pairs as $\{\theta^{(i)},\G(\theta^{(i)})\}_{i=1}^{M}$ and form the
design matrix $\mathsf{G} \in \mathbb{R}^{M\times d}$ in which the $i^{th}$ row
is the transpose of $\G(\theta^{(i)})$. In \citep{higdon2008}, it is suggested
to use PCA on the column space of $\mathsf{G}$ to determine new variables in
which to perform uncorrelated GP emulation. To this end, we average each of the
$d$ components of $\G(\theta^{(i)})$ over the $M$ training points to find the
mean output vector $m_{\mathsf{G}} \in \R^d$. Then, we form the design mean
matrix $\mathsf{M}_{\mathsf{G}} \in \R^{M \times d}$ by making each of its $M$
rows equal to the transpose of $m_{\mathsf{G}}$. We then perform an SVD to
obtain
\begin{align}
	(\mathsf{G} - \mathsf{M}_{\mathsf{G}}) = \hat{\mathsf{G}} D V^\top,
\end{align}
where $V \in \mathbb{R}^{d \times d}$ is orthogonal, $D \in \mathbb{R}^{d \times
d}$ is diagonal, and $\hat{\mathsf{G}}\in \mathbb{R}^{M \times d}$. The matrix
$\hat{\mathsf{G}}$ has orthogonal columns that represent uncorrelated output
coordinates. The matrix $D$ contains the unscaled standard deviations of the
original data $\mathsf{G}$. Lastly, $V$ contains the proportional loadings of
the original data coordinates \citep[see][]{jolliffe2011}. It is important to
note that the $i$-th row in $\mathsf{G}$ is related to the $i$-th row in
$\hat{\mathsf{G}}$, as both can be understood as the output of the $i$-th
ensemble member $\theta^{(i)}$ in our setting, albeit on an orthogonal
coordinate space.

We project the data onto the uncorrelated output space as $\hat y = D^{-1}V^\top
\, (y - m_{\mathsf{G}})$ and emulate using the resulting projections of the
model output as input-output training runs, $\{\theta^{(i)},D^{-1}V^\top \,
(\G(\theta^{(i)})-m_{\mathsf{G}})\}_{i=1}^{M}$, to obtain
\begin{align}\label{eq:pca_gp}
	\hat \G(\theta) \sim \N\left(\hat m(\theta), \hat \Gamma(\theta)\right).
\end{align}
Transforming back to the original output coordinates leads us to consider the
emulation of the forward model as
\begin{align}\label{eq:pca_gp_rec}
	\G(\theta) \sim \N \left( VD \, \hat m(\theta) + m_{\mathsf{G}}, VD\, \hat \Gamma(\theta)\, DV^\top\right),
\end{align}
This allows us to rewrite the misfit \eqref{eq:phi_gp} in the form of
\begin{align}
\Phigp(\theta;y) = \frac12 \|\hat y - \hat m(\theta) \|^2_{\hat \Gamma(\theta)} + \frac12 \log\det \hat{\Gamma(\theta)},
\end{align}
or compute either \eqref{eq:phi_m} or \eqref{eq:phi_gp2}, as discussed in
\cref{apx:time-diag}.

\end{document}